\documentclass[12pt]{article}

\usepackage{color}
\usepackage{array}
\usepackage{epsfig}
\usepackage{amssymb}
\usepackage{graphics,graphpap}
\usepackage{subfigure}
\usepackage{amsmath}

\newcommand{\bal}{\begin{align}}

\def\beq{\begin{equation}}
\def\eeq{\end{equation}}
\def\bea{\begin{eqnarray}}
\def\eea{\end{eqnarray}}
\def\nn{\nonumber}
\def\nl{\nonumber\\}

\def\roughly#1{\mathrel{\raise.3ex\hbox
{$#1$\kern-.75em\lower1ex\hbox{$\sim$}}}}

\def\lesssim{\mathrel{\hbox{\rlap{\hbox{\lower4pt\hbox{$\sim$}}}\hbox{$<$}}}}
\def\gtrsim{\mathrel{\hbox{\rlap{\hbox{\lower4pt\hbox{$\sim$}}}\hbox{$>$}}}}

\def\sla#1{\raise.15ex\hbox{$/$}\kern-.57em #1}

\def\ket#1{\left| #1\right\rangle}
\def\ks{K_S}

\newcommand{\ba}{\begin{array}}
\newcommand{\ea}{\end{array}}

\def\bd{B_d^0}
\def\bs{B_s^0}
\def\bdbar{{\bar B}_d^0}
\def\bsbar{{\bar B}_s^0}
\def\btod{{\bar b} \to {\bar d}}
\def\btos{{\bar b} \to {\bar s}}
\def\Bsdecay{\bs\to J/\psi \phi}
\def\Bsmumu{\bs \to \mu^+ \mu^-}
\def\bscc{{\bar b} \to {\bar s} c {\bar c}}
\def\bsss{{\bar b} \to {\bar s} s {\bar s}}

\def\bspp{\bs\to\phi\phi}

\allowdisplaybreaks
\begin{document}

\begin{flushright}  
UMISS-HEP-2012-06\\
UdeM-GPP-TH-12-208 \\
\end{flushright}

\begin{center}
\bigskip
{\Large \bf \boldmath New Physics in $\btos$ Transitions and the \\ $B_{d,s}^0 \to V_1 V_2$ Angular Analysis} \\

\bigskip
{\large Alakabha Datta $^{a,}$\footnote{datta@phy.olemiss.edu},
Murugeswaran Duraisamy $^{a,}$\footnote{duraism@phy.olemiss.edu} \\
and David London $^{b,}$\footnote{london@lps.umontreal.ca}}
\\
\end{center}

\begin{flushleft}
~~~~~~$a$: {\it Department of Physics and Astronomy, 108 Lewis Hall, }\\ 
~~~~~~~~~~{\it University of Mississippi, Oxford, MS 38677-1848, USA}\\
~~~~~~$b$: {\it Physique des Particules, Universit\'e de
Montr\'eal,}\\
~~~~~~~~~~{\it C.P. 6128, succ. centre-ville,
Montr\'eal, QC, Canada H3C 3J7}
\end{flushleft}
\begin{center} 
\vskip0.5cm
{\Large Abstract\\}
\vskip3truemm

\parbox[t]{\textwidth} {We suppose that there is new physics (NP) in
  $\btos$ transitions, and examine its effect on the angular
  distribution of $B^0_q \to V_1 V_2$ ($q=d,s$), where $V_{1,2}$ are
  vector mesons. We find that, in the presence of such NP, the
  formulae relating the parameters of the untagged, time-integrated
  angular distribution to certain observables (polarization fractions,
  CP-violating triple-product asymmetries, CP-conserving interference
  term) must be modified from their standard-model forms. This
  modification is due in part to a nonzero $B^0_q$-${\bar B}^0_q$
  width difference, which is significant only for $\bs$ decays. We
  re-analyze the $\bspp$ data to see the effect of these
  modifications. As $\Delta \Gamma_s/2 \Gamma_s \sim 10\%$, there are
  $O(10\%)$ changes in the derived observables.  These are not large,
  but may be important given that one is looking for signals of NP.
  In addition, if the NP contributes to the $\btos$ decay, the
  measurement of the untagged time-dependent angular distribution
  provides enough information to extract all the NP parameters.}

\end{center}


\baselineskip=14pt

\newpage

\section{Introduction}

As recently as a year ago, there were several hints of physics beyond
the standard model (SM) in $\btos$ transitions. For example, the CDF
\cite{CDF} and D\O\ \cite{D0} Collaborations measured the CP asymmetry
in $\Bsdecay$, and found a hint for indirect (mixing-induced) CP
violation. This is counter to the expectation of the SM, which
predicts this CP violation to be $\simeq 0$. In general, this result
was interpreted as evidence for a nonzero value of the weak phase of
$\bs$-$\bsbar$ mixing ($2\beta_s$), and the contributions of various
new-physics (NP) models to the $B_s$ mixing phase were explored
\cite{RPV,Z'FCNC, 2HDM, SUSY, littleHiggs,fourGen, Bspapers}.  It was
also pointed out that NP in the decay $\bscc$ could also play an
important role \cite{NPdecay}. In addition, the SM predicts that the
measured indirect CP asymmetry in $\bsss$ penguin decays should
generally be equal to that found in charmonium decays such as $\bd\to
J/\psi\ks$.  However, it was found that these two quantities were not
identical for several decays \cite{LunghiSoni}. As a third example,
the CDF Collaboration reported the measurement of $B(\Bsmumu) =
(1.8^{+1.1}_{-0.9}) \times 10^{-8}$ \cite{CDFmeas}.  This is larger
than the SM prediction for this branching ratio, which is $B(\Bsmumu)
= (3.35\pm 0.32)\times 10^{-9}$ \cite{blanke}. There were a number of
other effects -- in all cases, the disagreement with the SM was not
large, $\le 2\sigma$. Still, it was intriguing that all appear in
$\btos$ transitions.

However, with recent measurements, these effects have largely
disappeared, or at least been reduced. First, LHCb has measured the
indirect CP asymmetry in $\Bsdecay$, and finds $2\beta_s \simeq 0$, in
agreement with the SM \cite{LHCbBsmixing}. Specifically, they measure
$2\beta_s = (-0.06 \pm 5.77~({\rm stat}) \pm 1.54~({\rm
  syst}))^\circ$. Still, the errors are large enough that NP cannot be
excluded. Second, with the latest indirect CP asymmetry data, the
Heavy Flavor Averaging Group (HFAG) \cite{HFAG} finds that the
$\bd$-$\bdbar$ mixing phase $\sin 2\beta$ is measured to be (i) $0.68
\pm 0.02$ in charmonium decays, and (ii) $0.64 \pm 0.04$ in $\bsss$
penguin decays.  These numbers are quite similar, so that no real
discrepancy can be claimed.  On the other hand, several of the $\bsss$
decays have additional contributions with a different weak phase, and
so HFAG urges that the ``naive average'' in (ii) be used with extreme
caution.  Third, the recent LHCb update does not confirm the CDF
$\Bsmumu$ result \cite{LHCbupdate}. They improve the present upper
bound to $B(\Bsmumu) \le 1.3 \times 10^{-8}$ (90\% C.L.), in agreement
with the SM. Most of the other effects have similarly gone away, or
are simply not large enough to be compelling.

On the other hand, there is one discrepancy with the SM which has not
disappeared. The D\O\ Collaboration has reported an anomalously large
CP-violating like-sign dimuon charge asymmetry in the $B$ system. In
Ref.~\cite{D0dimuonold}, the asymmetry was found to be
\bea
  A_{\rm sl}^b = -(9.57 \pm 2.51 \pm 1.46) \times 10^{-3} ~,
\eea
which is a 3.2$\sigma$ deviation from the SM prediction, $A_{\rm
  sl}^{b,{\rm SM}} = ( -2.3^{+0.5}_{-0.6} ) \times
10^{-4}$ \cite{Lenz}. In fact, the updated measurement
\cite{D0dimuonnew} exhibits an even larger discrepancy:
\bea
  A_{\rm sl}^b = -(7.87 \pm 1.72 \pm 0.93) \times 10^{-3} ~,
\eea
a 3.9$\sigma$ deviation. Now, it has been shown that this anomaly is
due mainly to $B_s$ decays, i.e.\ a $\btos$ transition. So this is a
solid indication of $\btos$ NP.

Thus, at present the status of $\btos$ NP is uncertain. It seems
unlikely that the effect of such NP can be very large, but a smaller
effect is still possible. In this paper we make the assumption that NP
{\it is} present in $\btos$ transitions. However, in addition to
taking into account its effect on $B_s$ mixing, which is what is
conventionally done, we also consider its effect on $\btos$
decays. The main aim is to examine the effect of $\btos$ NP on the
angular distribution of $B^0_q \to V_1 V_2$ ($q=d,s$), where $V_{1,2}$
are vector mesons. In particular, we consider final states which are
self-conjugate, so that both $B^0_q$ and ${\bar B}^0_q$ can decay to
$V_1 V_2$, generating indirect (mixing-induced) CP-violating effects.

There are three classes of $B^0_q$ decays which can be affected by
$\btos$ NP:
\begin{enumerate}

\item $\bs$ decays with $\btos$,

\item $\bd$ decays with $\btos$,

\item $\bs$ decays with $\btod$.

\end{enumerate}
Our analysis is completely general and can be applied to any of these
classes. However, we also focus specifically on $\bspp$.  There are
two reasons. First, this is a pure $\btos$ penguin decay, and so there
can well be NP contributions to any of the loop-level penguin decay
amplitudes\footnote{$\bspp$ and $\Bsdecay$ were examined in
  Ref.~\cite{GRTP}, but only NP in $\bs$-$\bsbar$ mixing was
  considered, not NP in the decay.}. Second, the untagged angular
distribution of the decay has already been measured by the CDF
\cite{CDFBs} and LHCb \cite{LHCbphiphi} Collaborations, and so their
results can be (re)interpreted in the context of $\btos$ NP
contributions.

The result of this analysis -- and this is the main point of the paper
-- is as follows. The parameters of the untagged, time-integrated
angular distribution can be measured experimentally. Certain
observables can be derived from these parameters. However, in the
presence of NP, the formulae which relate the observables and
parameters are modified compared to their SM forms. There are six
terms ($i=1$-6) in the angular distribution, and we correspondingly
find six observables for which the relation between the experimental
data and theoretical parameters must be modified. For $i=1$-3 they are
the polarization fractions, for $i=4$,6 they are the CP-violating
triple-product asymmetries, and $i=5$ corresponds to a CP-conserving
observable. The modifications for the polarization fractions are
particularly striking. Here there are corrections to the SM formulae
that are proportional to the width difference in the $B^0_q$-${\bar
  B}^0_q$ system.  Now, the width difference $\Delta\Gamma$ is
sizeable only for $\bs$ decays. Thus, the formulae modifications due
to NP are important only for class-(1) and (3) decays, which include
$\bspp$. $\Delta \Gamma_s/2 \Gamma_s \sim 10\%$, so that the
modifications lead to $O(10\%)$ changes in the derived observables.
These are not large, but may be important given that one is looking
for signals of NP.

Another result is that, if the untagged, time-dependent angular
distribution can be measured, 12 observables can be obtained.  If the
NP contributes to the $\btos$ decay, there are fewer than 12 unknown
NP parameters. Thus, all of these parameters can be extracted from the
angular distribution. This may allow the identification of the NP.

We begin in Sec.~2 by presenting the most general $B^0_{d,s} \to V_1
V_2$ angular distribution, allowing for NP in the mixing and/or the
decay. We consider the angular distribution for several different
scenarios: at $t=0$ (Sec.~2.2.1), time-dependent (Sec.~2.2.2),
untagged time-dependent (Sec.~2.3), untagged time-integrated
(Sec.~2.4). In Sec.~3 we examine the untagged time-dependent and
time-integrated distributions for $\bspp$ within the SM. The study of
$\bspp$ is extended to the SM $+$ NP in Sec.~4. We discuss observables
such as the polarization fractions, CP-violating triple-product
asymmetries and the CP-conserving interference term, and note the
changes in the formulae used for their extraction necessitated by the
inclusion of $\btos$ NP. We also show that all the unknown NP
parameters in the $\btos$ decay can be determined from the measurement
of the untagged, time-dependent angular distribution.  In Sec.~5 we
present a numerical reanalysis of the $\bspp$ data allowing for the
possibility of $\btos$ NP contributions in the decay. We conclude in
Sec.~6.

\section{\boldmath $B \to V_1 V_2$ Angular Distribution}

\subsection{Generalities}

The most general Lorentz-covariant amplitude for the decay $B(p) \to
V_1(k_1,\varepsilon_1) + V_2(k_2,\varepsilon_2)$ is given by
\cite{Valencia,TPDL}
\beq
M = a \, \varepsilon_1^* \cdot \varepsilon_2^* + \frac{b}{m_B^2}
(p\cdot \varepsilon_1^*) (p\cdot \varepsilon_2^*) + i \frac{c}{m_B^2}
\epsilon_{\mu\nu\rho\sigma} p^\mu q^\nu \varepsilon_1^{*\rho}
\varepsilon_2^{*\sigma} ~,
\label{amp1}
\eeq
where $q\equiv k_1 - k_2$. The quantities $a$, $b$ and $c$ are complex
and contain in general both CP-conserving strong phases and
CP-violating weak phases.  In $B\to V_1 V_2$ decays, the final state
can have total spin 0, 1 or 2, which correspond to the $V_1$ and $V_2$
having relative orbital angular momentum $l=0$ ($s$ wave), $l=1$ ($p$
wave), or $l=2$ ($d$ wave), respectively.  The $a$ and $b$ terms
correspond to combinations of the parity-even $s$- and $d$-wave
amplitudes, while the $c$ term corresponds to the parity-odd $p$-wave
amplitude.

In order to obtain the angular distribution for $B\to V_1 V_2$, one
uses the linear polarization basis.  Here, one decomposes the decay
amplitude into components in which the polarizations of the
final-state vector mesons are either longitudinal ($A_0$), or
transverse to their directions of motion and parallel ($A_\|$) or
perpendicular ($A_\perp$) to one another. The transversity amplitudes
$A_h$ ($h=0,\|,\perp$) are related to $a$, $b$ and $c$ of
Eq.~(\ref{amp1}) via \cite{TPDL}
\beq
A_\| = \sqrt{2} a ~,~~~ A_0 = -a x - \frac{m_1 m_2}{m_B^2} b
(x^2 - 1) ~,~~~ A_\perp = 2\sqrt{2} \, \frac{m_1 m_2}{m_B^2} c 
\sqrt{x^2 - 1} ~,
\label{Aidefs}
\eeq
where $x = k_1 \cdot k_2 / (m_1 m_2)$ ($m_1$ and $m_2$ are the masses
of $V_1$ and $V_2$, respectively.).

The amplitude for ${\bar B}(p) \to {\bar V}_1(k_1,\varepsilon_1) +
{\bar V}_2(k_2,\varepsilon_2)$ can be obtained by operating on
Eq.~(\ref{amp1}) with CP. This yields
\beq
{\bar M} = {\bar a} \, \varepsilon_1^* \cdot \varepsilon_2^* + \frac{{\bar
    b}}{m_B^2} (p\cdot \varepsilon_1^*) (p\cdot \varepsilon_2^*)
- i \frac{{\bar c}}{m_B^2} \epsilon_{\mu\nu\rho\sigma} p^\mu q^\nu
\varepsilon_1^{*\rho} \varepsilon_2^{*\sigma} ~,
\eeq
in which ${\bar a}$, ${\bar b}$ and ${\bar c}$ are equal to $a$, $b$
and $c$, respectively, except that the weak phases are of opposite
sign. The above equation can be obtained from Eq.~(\ref{amp1}) by
changing $a \to {\bar a}$, $b \to {\bar b}$ and $c \to -{\bar c}$.
Similarly, one defines ${\bar A}_0$, ${\bar A}_\|$, and ${\bar
  A}_\perp$, which are equal to $A_0$, $A_\|$, and $A_\perp$,
respectively, but with weak phases of opposite sign.

\subsection{\boldmath $B^0_{d,s} \to V_1 V_2$}
\label{VVangdist}

As mentioned in the introduction, we are interested in the decays
$B^0_q \to V_1 V_2$ ($q=d,s$), in which both $B^0_q$ and ${\bar
  B}^0_q$ can decay to $V_1 V_2$. Due to $B^0_q$-${\bar B}^0_q$
mixing, the amplitude is time dependent. Assuming that $V_{1,2}$ both
decay into two pseudoscalars, i.e.\ $V_1 \to P_1 P_1'$, $V_2 \to P_2
P_2'$, the angular distribution is given in terms of the vector
$\vec{\omega} = (\cos{\theta_1 },\cos{\theta_2 }, \Phi)$
\cite{DDLR,CW}:
\bea
\label{AD:Bsphiphi}
\frac{d^4 \Gamma(t)}{dt d\vec{\omega}} &=& \frac{9}{32 \pi} \sum^6_{i=1} K_i(t) f_i(\vec{\omega}) ~.
\eea
Here, $\theta_1$ ($\theta_2$) is the angle between the directions of
motion of the $P_1$ ($P_2$) in the $V_1$ ($V_2$) rest frame and the
$V_1$ ($V_2$) in the $B$ rest frame, and $\Phi$ is the angle between
the normals to the planes defined by $P_1 P_1'$ and $P_2 P_2'$ in the
$B$ rest frame. The angular-dependent  terms are given by
\bea
f_1(\vec{\omega}) = 4\cos^2{\theta_1} \cos^2{\theta_2} & ~~,~~ &
f_2 (\vec{\omega})= 2 \sin^2{\theta_1} \sin^2{\theta_2} \cos^2{\Phi} ~, \nl
f_3(\vec{\omega}) = 2 \sin^2{\theta_1} \sin^2{\theta_2} \sin^2{\Phi} & ~~,~~ &
f_4(\vec{\omega}) = -2\sin^2{\theta_1}\sin^2{\theta_2}  \sin{2\Phi} ~, \nl
f_5(\vec{\omega}) =\sqrt{2}\sin{2\theta_1} \sin{2\theta_2} \cos{\Phi} & ~~,~~ &
f_6(\vec{\omega}) = - \sqrt{2}\sin{2\theta_1} \sin{2\theta_2} \sin{\Phi}~. 
\eea

\subsubsection{\boldmath $t=0$}

At $t=0$, the $K_i$ are
\bea
\label{Kis:Bsphiphi}
& K_1 = |A_0|^2 ~~,~~
K_2 =  |A_\parallel|^2  ~~,~~
K_3 =  |A_\perp|^2  ~, & \nl
& K_4 = {\rm Im}(A_\perp A^*_\parallel) ~~,~~
K_5 = {\rm Re}(A_\parallel  A^*_{0}) ~~,~~
K_6 = {\rm Im}(A_\perp A^*_{0}) ~. &
\eea
The angular distribution for the CP-conjugate decay $\bar{B}^0_q
\to V_1 V_2$ is the same as that given above with the replacements
$K_i \to {\bar K}_i$, and $A_h \to {\bar A}_h$.

The quantities $K_4$ and $K_6$ are particularly interesting. They are
related to the $\epsilon_{\mu\nu\rho\sigma} p^\mu q^\nu
\varepsilon_1^{*\rho} \varepsilon_2^{*\sigma}$ term of
Eq.~(\ref{amp1}), which is proportional to ${\vec q} \cdot
({\vec\varepsilon}_1 \times {\vec\varepsilon}_2)$ in the rest frame of
the $B$. This is the triple product (TP).  The TP is odd under both
parity and time reversal, and thus constitutes a potential signal of
CP violation. However, here one has to be a bit careful. As noted
above, the $A_h$ possess both weak (CP-odd) and
strong (CP-even) phases. Thus, $K_4$ and/or $K_6$ can be nonzero even
if the weak phases vanish. In order to obtain a true signal of CP
violation, one has to compare the $B$ and ${\bar B}$ decays.  Now,
${\bar K}_4$ is the same as $K_4$, except that (i) the weak phases
change sign, and (ii) there is an overall relative minus sign due to
the presence of ${\bar A}_\perp$/$A_\perp$, and similarly for ${\bar
  K}_6$ and $K_6$. This implies that the true (CP-violating) TP's are
given by the untagged observables $K_4 + {\bar K}_4$ and $K_6 + {\bar
  K}_6$.  There are also fake (CP-conserving) TP's, due only to strong
phases of the the $A_h$'s, given by $K_4 - {\bar K}_4$ and $K_6 -
{\bar K}_6$.

\subsubsection{Time dependence}

In order to calculate the $K_i(t)$, one proceeds as follows. Due to
$B^0_q$-${\bar B}^0_q$ mixing, the time evolution of the states
$\ket{B^0_q (t)}$ and $\ket{{\bar B}^0_q(t)}$ can be described by the
relations \cite{Anikeev:2001rk}
\bea
\label{Timeev:Bst}
\ket{B^0_q (t)} &=& g_+(t) \,\ket{B^0_q}  + \frac{q}{p}\, g_-(t) \, \ket{{\bar B}_q^0}   ~, \nl
\ket{{\bar B}_q^0 (t)} &=& \frac{p}{q}\, g_-(t) \,\ket{B_q^0}  + \, g_+(t) \, \ket{{\bar B}_q^0}   ~,
\eea 
where $q/p = e^{-i\phi_q}$. Here, $\phi_q$ is the phase of
$B^0_{q}$-${\bar B}^0_{q}$ mixing. In the SM, we have $\phi_d = 2\beta
= (42.8 \pm 1.6)^\circ$ from charmonium decays \cite{HFAG}. Also,
assuming no NP in the decay, the LHCb Collaboration measures $\phi_s =
(-0.06 \pm 5.77~({\rm stat}) \pm 1.54~({\rm syst}))^\circ$ in $\bs\to
J/\psi \phi$ \cite{LHCbBsmixing}. Although this agrees with the SM
prediction of $\phi_s \simeq 0$, the errors are still large enough
that NP in the decay and/or mixing cannot be excluded.

In the above, we have
\bea
g_+(t) &=& \frac12 \Big(e^{-(i M_{L} + \Gamma_{L}/2)t} \,+ e^{-(i M_{H} + \Gamma_{H}/2)t}\Big)  ~, \nl
g_-(t) &=& \frac12  \Big(e^{-(i M_{L} + \Gamma_{L}/2)t} \,- e^{-(i M_{H} + \Gamma_{H}/2)t}\Big) ~,
\eea
where $L$ and $H$ indicate the light and heavy states, respectively.
The average mass and width, and the mass and width differences of the
$B$-meson eigenstates are defined by
\begin{equation}
\begin{array}{rclrcl}
m & = & \displaystyle \frac{ M_H + M_L}{2}   ~, \qquad  \qquad  & 
  \Gamma & = & \displaystyle \frac{\Gamma_L + \Gamma_H}2   ~,
  \\[6pt] 
\Delta m & = & M_H - M_L  ~, & \Delta \Gamma & = & \Gamma_L - \Gamma_H ~.
\end{array}\label{mg} 
\end{equation}
$\Delta m$ is positive by definition. For $\bd$ mesons, $\Gamma_L
\simeq \Gamma_H$, so that $\Delta \Gamma_d = 0$. However, for $\bs$
mesons, $\Delta \Gamma_s$ is reasonably large: $|\Delta \Gamma_s| =
0.123 \pm 0.029~({\rm stat}) \pm 0.011~({\rm syst})~{\rm ps}^{-1}$
\cite{LHCb}.  In our convention the SM prediction for $\Delta
\Gamma_s$ is positive, and it has been recently confirmed
experimentally that $\Delta \Gamma_s > 0$ \cite{DelGam2meas}.

The time dependence of the transversity amplitudes $A_h$ is due to
$B^0_q$-${\bar B}^0_q$ mixing. For the decay to a final state $f$ we
have
\bea
\label{Tdep:TransAmp}
A_h(t) &=&  \langle f | B^0_q (t) \rangle_h = \Big[ g_+(t) A_h+ \eta_h ~ q/p ~g_-(t)~ \bar{A}_h\Big] ~, \nl 
\bar{A}_h(t) &=&  \langle f | \bar{B}^0_q (t) \rangle_h = \Big[ p/q~g_-(t) A_h+ \eta_h ~ g_+(t)~ \bar{A}_h\Big] ~,
\eea
where $A_h = \langle f | B^0_q \rangle_h$, $\bar{A}_h = \langle f
|\bar{B}^0_q \rangle_h$, and $\eta_{0, \parallel} = 1$, $\eta_\perp =
-1$. In calculating the $K_i(t)$, the following relations are useful:
\bea
\label{gplusminus}
|g_\pm(t)|^2 &=& \frac12 e^{ - \Gamma t} \Big(\cosh{(\Delta \Gamma/2) t}\,\pm \cos{ \Delta m t}\Big)  ~, \nl
g^*_+(t) g_-(t) &=& \frac12 e^{ - \Gamma t} \Big(-\sinh{(\Delta \Gamma/2) t}\,+ i  \sin{ \Delta m t}\Big) ~.
\eea

The expressions for the time-dependent functions $K_i(t)$ are given by
\bea
\label{Kits}
K_1(t) = |A_0(t)|^2 & = & \frac12 e^{ - \Gamma t} \left[ \left(
  |A_0|^2 + |{\bar A}_0|^2 \right) \cosh{(\Delta \Gamma/2) t} \right. \nl
&& \hskip0.7truecm  +~\left(
  |A_0|^2 - |{\bar A}_0|^2 \right) \cos{ \Delta m t} \nl
&& \hskip-0.7truecm -~2{\rm Re}(A_0^* \bar{A}_0) \left( \cos \phi_q \sinh{(\Delta
    \Gamma/2) t} - \sin \phi_q \sin{ \Delta m t} \right) \nl
&& \hskip-0.7truecm \left.  -~2{\rm Im}(A_0^* \bar{A}_0) \left( \cos \phi_q \sin{ \Delta
    m t} + \sin \phi_q \sinh{(\Delta \Gamma/2) t} \right) \right] ~, \nl
K_2(t) = |A_\|(t)|^2 & = & \frac12 e^{ - \Gamma t} \left[ \left(
  |A_\||^2 + |{\bar A}_\||^2 \right) \cosh{(\Delta \Gamma/2) t} \right. \nl
&& \hskip0.7truecm +~\left(
  |A_\||^2 - |{\bar A}_\||^2 \right) \cos{ \Delta m t} \nl
&& \hskip-0.7truecm -~2{\rm Re}(A_\|^* \bar{A}_\|) \left( \cos \phi_q \sinh{(\Delta
    \Gamma/2) t} - \sin \phi_q \sin{ \Delta m t} \right) \nl
&& \hskip-0.7truecm \left.  -~2{\rm Im}(A_\|^* \bar{A}_\|) \left( \cos \phi_q \sin{ \Delta
    m t} + \sin \phi_q \sinh{(\Delta \Gamma/2) t} \right) \right] ~, \nl
K_3(t) = |A_\perp(t)|^2 & = & \frac12 e^{ - \Gamma t} \left[ \left(
  |A_\perp|^2 + |{\bar A}_\perp|^2 \right) \cosh{(\Delta \Gamma/2) t} \right. \nl
&& \hskip0.7truecm +~\left(
  |A_\perp|^2 - |{\bar A}_\perp|^2 \right) \cos{ \Delta m t} \nl
&& \hskip-0.7truecm +~2{\rm Re}(A_\perp^* \bar{A}_\perp) \left( \cos \phi_q \sinh{(\Delta
    \Gamma/2) t} - \sin \phi_q \sin{ \Delta m t} \right) \nl
&& \hskip-0.7truecm \left.  +~2{\rm Im}(A_\perp^* \bar{A}_\perp) \left( \cos \phi_q \sin{ \Delta
    m t} + \sin \phi_q \sinh{(\Delta \Gamma/2) t} \right) \right] ~, \nl
K_4(t) = {\rm Im} ( A_\perp(t) A_\|^*(t) ) & = & \frac12 e^{ - \Gamma t} \left[ 
     \left( {\rm Im} ( A_\perp A_\|^* ) - {\rm Im} ( \bar{A}_\perp \bar{A}_\|^* ) \right) \cosh{(\Delta \Gamma/2) t} \right. \nl
&& \hskip0.7truecm +~\left( {\rm Im} ( A_\perp A_\|^* ) + {\rm Im} ( \bar{A}_\perp \bar{A}_\|^* ) \right) \cos{ \Delta m t} \nl
&& \hskip-2.8truecm
      +~ \left( {\rm Im} ( A_\perp \bar{A}_\|^* ) - {\rm Im} ( \bar{A}_\perp A_\|^* ) \right)
\left( -\sinh{(\Delta \Gamma/2) t} \cos \phi_q + \sin{ \Delta m t} \sin \phi_q \right) \nl
&& \hskip-2.8truecm \left. 
            +~ \left( {\rm Re} ( A_\perp \bar{A}_\|^* ) + {\rm Re} ( \bar{A}_\perp A_\|^* ) \right) 
\left( -\sinh{(\Delta \Gamma/2) t} \sin \phi_q - \sin{ \Delta m t} \cos \phi_q \right) \right] ~, \nl
K_5(t) = {\rm Re} ( A_\|(t) A_0^*(t) ) & = & \frac12 e^{ - \Gamma t} \left[ 
        \left( {\rm Re}(A_\| A_0^*) + {\rm Re}(\bar{A}_\| \bar{A}_0^*) \right) \cosh{(\Delta \Gamma/2) t} \right. \nl
&& \hskip0.7truecm +~\left( {\rm Re}(A_\| A_0^*) - {\rm Re}(\bar{A}_\| \bar{A}_0^*) \right) \cos{ \Delta m t} \nl
&& \hskip-2.8truecm
           +~ \left( {\rm Re}(A_\| \bar{A}_0^*) + {\rm Re}(\bar{A}_\| A_0^*) \right) 
               \left( -\sinh{(\Delta \Gamma/2) t} \cos \phi_q + \sin{ \Delta m t} \sin \phi_q \right) \nl
&& \hskip-2.8truecm \left.
           +~ \left( {\rm Im}(A_\| \bar{A}_0^*) - {\rm Im}(\bar{A}_\| A_0^*) \right) 
               \left( \sinh{(\Delta \Gamma/2) t} \sin \phi_q + \sin{ \Delta m t} \cos \phi_q \right) \right] ~, \nl
K_6(t) = {\rm Im} ( A_\perp(t) A_0^*(t) ) & = & \frac12 e^{ - \Gamma t} \left[ 
     \left( {\rm Im} ( A_\perp A_0^* ) - {\rm Im} ( \bar{A}_\perp \bar{A}_0^* ) \right) \cosh{(\Delta \Gamma/2) t} \right. \nl
&& \hskip0.7truecm +~\left( {\rm Im} ( A_\perp A_0^* ) + {\rm Im} ( \bar{A}_\perp \bar{A}_0^* ) \right) \cos{ \Delta m t} \\
&& \hskip-2.8truecm
      +~ \left( {\rm Im} ( A_\perp \bar{A}_0^* ) - {\rm Im} ( \bar{A}_\perp A_0^* ) \right)
\left( -\sinh{(\Delta \Gamma/2) t} \cos \phi_q + \sin{ \Delta m t} \sin \phi_q \right) \nl
&& \hskip-2.8truecm \left. 
            +~ \left( {\rm Re} ( A_\perp \bar{A}_0^* ) + {\rm Re} ( \bar{A}_\perp A_0^* ) \right) 
\left( -\sinh{(\Delta \Gamma/2) t} \sin \phi_q - \sin{ \Delta m t} \cos \phi_q \right) \right] ~. \nn
\eea

The expressions for the time-dependent $\bar{K}_i(t)$'s can be
obtained from the $K_i(t)$'s by changing the sign of the weak phases in
both the decay ($A_h \leftrightarrow \eta_h {\bar A}_h$) and the mixing
($\phi_q \to -\phi_q$).

\subsection{Untagged decays}
\label{untagged}

In the previous subsections, we presented the angular distribution for
the case in which the initial decay meson is tagged, so that one can
distinguish the $B^0_q$ and ${\bar B}^0_q$ decays. In practice,
however, tagging is difficult. Thus, as a first step, experiments will
examine the untagged decay, and this is considered here.

The untagged time-dependent angular distribution is given by
\bea
\frac{d^4 (\Gamma^{B_q} + \Gamma^{\bar{B}_q})}{dt d\vec{\omega}} &=& \frac{9}{32 \pi} \sum^6_{i=1} (K_i(t) + \bar{K}_i(t) ) f_i(\vec{\omega})~,
\eea
where the untagged observables can be found using Eq.~(\ref{Kits}):
\bea
\label{UTKis}
K_1(t)+\bar{K}_1(t) &=&  e^{ - \Gamma t} \Big[ \Big(
  |A_0|^2 + |{\bar A}_0|^2 \Big) \cosh{(\Delta \Gamma/2) t}    \nl 
&& \hskip-1.0truecm 
-~2\Big({\rm Re}(A_0^* \bar{A}_0) \cos \phi_q + {\rm Im}(A_0^* \bar{A}_0) \sin \phi_q \Big)\sinh{(\Delta \Gamma/2) t} \Big] ~, \nl
K_2(t)+\bar{K}_2(t) &=&  e^{ - \Gamma t}  \Big[ \Big(
  |A_\||^2 + |{\bar A}_\||^2 \Big) \cosh{(\Delta \Gamma/2) t}  \nl 
&& \hskip-1.0truecm 
-~2 \Big({\rm Re}(A_\|^* \bar{A}_\|) \cos \phi_q  + {\rm Im}(A_\|^* \bar{A}_\|) \sin \phi_q \Big) \sinh{(\Delta \Gamma/2) t} \Big] ~, \nl
K_3(t)+\bar{K}_3(t) &=&  e^{ - \Gamma t} \Big[ \Big(
  |A_\perp|^2 + |{\bar A}_\perp|^2 \Big)  \cosh{(\Delta \Gamma/2) t} \nl 
&& \hskip-1.0truecm  
+~2 \Big({\rm Re}(A_\perp^* \bar{A}_\perp) \cos \phi_q + {\rm Im}(A_\perp^* \bar{A}_\perp) \sin \phi_q \Big) \sinh{(\Delta \Gamma/2) t} \Big] ~, \nl
K_4(t) +\bar{K}_4(t) &=&  e^{ - \Gamma t}  \Big[ \Big( {\rm Im} ( A_\perp A_\|^* ) - {\rm Im} ( \bar{A}_\perp \bar{A}_\|^* ) \Big) \cosh{(\Delta \Gamma/2) t} \nl 
&&  -~ \Big(( {\rm Im} ( A_\perp \bar{A}_\|^* ) - {\rm Im} ( \bar{A}_\perp A_\|^* ) )\cos \phi_q \nl 
&& +~( {\rm Re} ( A_\perp \bar{A}_\|^* ) + {\rm Re} ( \bar{A}_\perp A_\|^* ) ) \sin \phi_q\Big)
\sinh{(\Delta \Gamma/2) t}  \Big] ~, \nl
K_5(t)+\bar{K}_5(t) &=& e^{ - \Gamma t}  \Big[ \Big( {\rm Re}(A_\| A_0^*) + {\rm Re}(\bar{A}_\| \bar{A}_0^*) \Big) \cosh{(\Delta \Gamma/2) t} \nl 
&&      -~\Big( ( {\rm Re}(A_\| \bar{A}_0^*) + {\rm Re}(\bar{A}_\| A_0^*) ) 
               \cos \phi_q  \nl 
&&          -~( {\rm Im}(A_\| \bar{A}_0^*) - {\rm Im}(\bar{A}_\| A_0^*) ] 
              \sin \phi_q  \Big) \sinh{(\Delta \Gamma/2) t} \Big] ~, \nl
K_6(t) +\bar{K}_6(t) &=&  e^{ - \Gamma t}  \Big[ \Big( {\rm Im} ( A_\perp A_0^* ) - {\rm Im} ( \bar{A}_\perp \bar{A}_0^* ) \Big) 
\cosh{(\Delta \Gamma/2) t} \nl 
&&  -~ \Big(( {\rm Im} ( A_\perp \bar{A}_0^* ) - {\rm Im} ( \bar{A}_\perp A_0^* ) )\cos \phi_q \nl 
&& +~( {\rm Re} ( A_\perp \bar{A}_0^* ) + {\rm Re} ( \bar{A}_\perp A_0^* ) ) \sin \phi_q\Big)
\sinh{(\Delta \Gamma/2) t}  \Big] ~.
\eea
Note that the CP properties of all the terms are respected. For
example, the $K_i(t) +\bar{K}_i(t)$ ($i=1,2,3,5$) are supposed to be
CP-even. But they contain terms proportional to $\sin \phi_q$, which
is CP-odd. This is accounted for because, in all cases, $\sin \phi_q$
is multipled by a term involving the helicity amplitudes which is also
CP-odd. Similarly, $\cos \phi_q$ (CP-even) is multipled by a
helicity-amplitude term which is also CP-even. The upshot is that
the $K_i(t) +\bar{K}_i(t)$ ($i=1,2,3,5$) are indeed CP-even. And it is
straightforward to verify that the $K_i(t) +\bar{K}_i(t)$ ($i=4,6$) are
CP-odd.

The key point here is the following. The individual $K_i$'s and ${\bar
  K}_i$'s [Eq.~(\ref{Kits})] depend on four functions of time: $e^{ -
  \Gamma t}\cos{ \Delta m t}$, $e^{ - \Gamma t}\sin{ \Delta m t}$,
$e^{ - \Gamma t}\cosh{(\Delta \Gamma/2) t}$, and $e^{ - \Gamma
  t}\sinh{(\Delta \Gamma/2) t}$. However, in the expressions above,
the dependence on the functions $e^{ - \Gamma t}\cos{ \Delta m t}$ and
$e^{ - \Gamma t}\sin{ \Delta m t}$ cancels, so that the untagged
observables depend only on $e^{ - \Gamma t}\cosh{(\Delta \Gamma/2) t}$
and $e^{ - \Gamma t}\sinh{(\Delta \Gamma/2) t}$. For $\bd$ mesons,
$\Delta \Gamma = 0$, so that the untagged observables are equal to
$e^{ - \Gamma t} \times$ simple sums of functions of the $A_i$ and
${\bar A}_i$. On the other hand, since $\Delta \Gamma \ne 0$ for $\bs$
mesons, the untagged observables are now complicated functions of the
$A_i$ and ${\bar A}_i$.

In addition, we have
\beq
\label{expLR}
e^{ - \Gamma t} \cosh{(\Delta \Gamma/2) t} = \frac12 \left( e^{ - \Gamma_L t} + e^{ - \Gamma_H t} \right) ~~,~~~~
e^{ - \Gamma t} \sinh{(\Delta \Gamma/2) t} = \frac12 \left( e^{ - \Gamma_L t} - e^{ - \Gamma_H t} \right) ~.
\eeq
If the $e^{-\Gamma_L t/2}$ and $e^{-\Gamma_H t/2}$ terms can be
distinguished experimentally, which is doable for $\bs$ decays, the
untagged time-dependent angular distribution provides 12 observables,
2 for each $K_i(t) +\bar{K}_i(t)$. Thus, $\bs \to V_1 V_2$ decays are
particularly interesting.

\subsection{Time-integrated untagged distribution}

As noted in the previous subsection, because $\Delta \Gamma \ne 0$ for
$\bs$ mesons, $\bs$ decays can be treated without tagging. The
time-integrated untagged angular distribution can be obtained by
integrating the $K_i(t)+\bar{K}_i(t)$ observables over time:
\bea
\label{ADphiphi-inte}
\frac{d^3 \langle\Gamma(\bs \to f) \rangle}{ d\vec{\omega} } &=&
\frac{9}{32 \pi} \sum^6_{i=1} \langle K_i \rangle f_i(\vec{\omega})~,
\eea
where
\bea
\label{TIKis}
\langle\Gamma(\bs \to f) \rangle   &=& \frac{1}{2}\int^\infty_0 dt (\Gamma^{B_s} + \Gamma^{\bar{B}_s})~,\nl
\langle K_i \rangle &=& \frac{1}{2}\int^\infty_0 dt(K_i(t)+\bar{K}_i(t))~.
\eea

One can obtain the $\langle K_i \rangle$'s from Eq.~(\ref{UTKis}):
\bea
\label{UTKisTI}
\langle K_1\rangle &=&  \frac{\tau_{B_s}}{2(1-y^2_s)} \Big[ \Big(
  |A_0|^2 + |{\bar A}_0|^2 \Big)   -2\Big({\rm Re}(A_0^* \bar{A}_0) \cos \phi_s +~{\rm Im}(A_0^* \bar{A}_0) \sin \phi_s \Big)y_s \Big] ~, \nl
\langle K_2\rangle &=&  \frac{\tau_{B_s}}{2(1-y^2_s)} \Big[ \Big(
  |A_\||^2 + |{\bar A}_\||^2 \Big)   -2\Big({\rm Re}(A_\|^* \bar{A}_\|) \cos \phi_s +~{\rm Im}(A_\|^* \bar{A}_\|) \sin \phi_s \Big)y_s \Big] ~, \nl
\langle K_3\rangle &=&  \frac{\tau_{B_s}}{2(1-y^2_s)} \Big[ \Big(
  |A_\perp|^2 + |{\bar A}_\perp|^2 \Big)+ 2 \Big({\rm Re}(A_\perp^* \bar{A}_\perp) \cos \phi_s  + ~ {\rm Im}(A_\perp^* \bar{A}_\perp) \sin \phi_s \Big) y_s \Big] ~, \nl
\langle K_4\rangle &=&  \frac{\tau_{B_s}}{2(1-y^2_s)}   \Big[ \Big( {\rm Im} ( A_\perp A_\|^* ) - {\rm Im} ( \bar{A}_\perp \bar{A}_\|^* ) \Big) -~ \Big(( {\rm Im} ( A_\perp \bar{A}_\|^* ) - {\rm Im} ( \bar{A}_\perp A_\|^* ) )\cos \phi_s \nl 
&& \hskip2.5truecm  
+~( {\rm Re} ( A_\perp \bar{A}_\|^* ) + {\rm Re} ( \bar{A}_\perp A_\|^* ) ) \sin \phi_s\Big)
y_s \Big] ~, \nl
\langle K_5\rangle &=& \frac{\tau_{B_s}}{2(1-y^2_s)}  \Big[ \Big( {\rm Re}(A_\| A_0^*) + {\rm Re}(\bar{A}_\| \bar{A}_0^*) \Big) 
          -~\Big( ( {\rm Re}(A_\| \bar{A}_0^*) + {\rm Re}(\bar{A}_\| A_0^*) ) 
               \cos \phi_s  \nl 
&& \hskip2.5truecm 
         -~( {\rm Im}(A_\| \bar{A}_0^*) - {\rm Im}(\bar{A}_\| A_0^*) ) 
              \sin \phi_s  \Big) y_s \Big] ~, \nl 
\langle K_6\rangle &=& \frac{\tau_{B_s}}{2(1-y^2_s)} \Big[ \Big( {\rm Im} ( A_\perp A_0^* ) - {\rm Im} ( \bar{A}_\perp \bar{A}_0^* ) \Big) -~ \Big(( {\rm Im} ( A_\perp \bar{A}_0^* ) - {\rm Im} ( \bar{A}_\perp A_0^* ) )\cos \phi_s \nl 
&& \hskip2.5truecm 
+~( {\rm Re} ( A_\perp \bar{A}_0^* ) + {\rm Re} ( \bar{A}_\perp A_0^* ) ) \sin \phi_s\Big)
y_s  \Big] ~,
\eea 
where $y_s \equiv \Delta \Gamma_s/2 \Gamma_s$.

At this stage, one clearly sees the effect of a nonzero $\Delta
\Gamma_s$ (or $y_s$). For $\bd$ decays, $\Delta \Gamma_d = 0$, so
there are no terms proportional to $y_d \equiv \Delta \Gamma_d/2
\Gamma_d$ in the $\langle K_i\rangle$. Indeed, the $\langle
K_i\rangle$ take the same form as the
$(K_i(t)+\bar{K}_i(t))\vert_{t=0}$ [Eq.~(\ref{Kis:Bsphiphi})].
However, this does not hold for $\bs$ decays. Because of the nonzero
$y_s$, the $\langle K_i\rangle$, which are time-integrated quantities,
take a different form than they did at $t=0$. And this means that, if
general $\btos$ NP is considered, the formulae relating certain
observables to the $\langle K_i\rangle$ must necessarily include terms
proportional to $y_s$. As we will see, this holds specifically for the
polarization fractions, CP-violating triple-product asymmetries, and
the CP-conserving interference term.

\subsection{Effective lifetime}

In general, the expressions for $K_i(t)+\bar{K}_i(t)$
[Eq.~(\ref{UTKis})] and $\langle K_i \rangle$ [Eq.~(\ref{UTKisTI})] have
the form
\bea
\label{effT1}
 K_i(t)+\bar{K}_i(t) &=& 2e^{ - \Gamma t} \Big[ {\cal{A}}^{ch}_i  \cosh{(\Delta \Gamma/2) t} + {\cal{A}}^{sh}_i \sinh{(\Delta \Gamma/2) t} \Big] ~, \nl
 \langle K_i \rangle &=&  \frac{\tau_{B_s}}{(1-y^2_s)}  \Big[ {\cal{A}}^{ch}_i  + {\cal{A}}^{sh}_i y_s \Big] ~,
\eea
where the experimental observables (dependent on the $K_i$) are on the
left-hand side, and the theoretical expressions (dependent on
${\cal{A}}^{ch}_i $ and ${\cal{A}}^{sh}_i $) are on the right-hand
side. (We have implicitly assumed that $\Delta \Gamma \ne 0$, which
implies a $\bs$ decay.)  ${\cal{A}}^{ch}_i $ and ${\cal{A}}^{sh}_i $
can be related to the experimental observables via the effective
lifetime \cite{fleischer}:
\bea
\label{effT2}
\tau^{eff, i}_{B_s} &=& \frac{\int^\infty_0 t (K_i(t)+\bar{K}_i(t))dt}{\int^\infty_0  (K_i(t)+\bar{K}_i(t))dt}\, \nl
&=&  \frac{\tau_{B_s}}{(1 - y_s^2)}\frac{( 1 + 
    2 {\cal{A}}^{i}_{\Delta \Gamma} y_s + y_s^2)}{(1 +    {\cal{A}}^{i}_{\Delta \Gamma} y_s)} ~,
\eea
where ${\cal{A}}^{i}_{\Delta \Gamma} \equiv {\cal{A}}^{sh}_i/{\cal{A}}^{ch}_i $.

Using Eqs.~(\ref{effT1}) and (\ref{effT2}), one can relate the
${\cal{A}}^{ch}_i $ to the $\langle K_i \rangle$:
\bea
\label{effT3}
{\cal{A}}^{ch}_i &=& \frac{\langle K_i \rangle}{\tau_{B_s}}  \Big(2 -\frac{\tau^{eff, i}_{B_s}}{\tau_{B_s}}  (1 - y_s^2) \Big)~.
\eea
The ${\cal{A}}^{sh}_i$ can be obtained from ${\cal{A}}^{i}_{\Delta \Gamma}$.

\section{\boldmath $\bspp$ -- SM}

The results of the previous section are completely general. In this
section we focus on the angular distribution of the pure $\btos$
penguin decay $\bspp$ within the SM.

In the SM, the amplitude for $\bspp$ can be written
\bea
{\cal A}(\bspp) &=& \lambda^{(s)}_t P'_t + \lambda^{(s)}_c P'_c + \lambda^{(s)}_u P'_u \nn\\
         &=& \lambda^{(s)}_t P'_{tc} + \lambda^{(s)}_u P'_{uc} ~,
\label{Bsampcelim}
\eea
where $\lambda^{(s)}_q \equiv V_{qb}^* V_{qs}$. (As this is a $\btos$
transition, the diagrams are written with primes.) In the second line,
we have used the unitarity of the Cabibbo-Kobayashi-Maskawa (CKM)
matrix ($\lambda^{(s)}_u + \lambda^{(s)}_c + \lambda^{(s)}_t = 0$) to
eliminate the $c$-quark contribution: $P'_{tc} \equiv P'_t - P'_c$,
$P'_{uc} \equiv P'_u - P'_c$.

Now, $|\lambda^{(s)}_t|$ and $|\lambda^{(s)}_u|$ are $O(\lambda^2)$
and $O(\lambda^4)$, respectively, where $\lambda=0.23$ is the sine of
the Cabibbo angle.  This suggests that the $\lambda^{(s)}_u P'_{uc}$
term can be neglected compared to $\lambda^{(s)}_t P'_{tc}$. However,
if one does this, one must be consistent and neglect {\it all}
$O(\lambda^4)$ terms. In particular, ${\rm Im}(\lambda^{(s)}_t)$ is
$O(\lambda^4)$, and so it too can be neglected. But since $2\beta_s =
-\arg \left( (q/p) ({\bar{\cal A}}/{\cal A}) \right)$, one also has
$\beta_s = 0$ because $(q/p) = ({\bar{\cal A}}/{\cal A}) = 1$ in the
limit where $\lambda^{(s)}_t$ is real. Thus, in the approximation of
neglecting all quantities of $O(\lambda^4)$, there are no nonzero weak
phases in $\bspp$, either in the mixing or in the decay.

\subsection{Untagged distribution}

In the approximation of neglecting all weak phases in $\bspp$, the
untagged observables [Eq.~(\ref{UTKis})] are
\bea
\label{UTKisSM}
(K_1(t)+\bar{K}_1(t))_{SM} &=&  e^{ - \Gamma t} \Big[ 2
  |A_0|^2 \Big ( \cosh{(\Delta \Gamma/2) t}  -\sinh{(\Delta \Gamma/2) t} \Big) \Big] ~, \nl
(K_2(t)+\bar{K}_2(t))_{SM} &=&  e^{ - \Gamma t} \Big[ 2
  |A_\||^2 \Big ( \cosh{(\Delta \Gamma/2) t}  -\sinh{(\Delta \Gamma/2) t} \Big) \Big] ~, \nl
(K_3(t)+\bar{K}_3(t))_{SM} &=&  e^{ - \Gamma t} \Big[ 2
  |A_\perp|^2 \Big ( \cosh{(\Delta \Gamma/2) t}  +\sinh{(\Delta \Gamma/2) t} \Big) \Big] ~, \nl
(K_4(t) +\bar{K}_4(t))_{SM} &=& 0 ~, \nl
(K_5(t)+\bar{K}_5(t))_{SM} &=& e^{ - \Gamma t}  \Big[  2 {\rm Re}(A_\| A_0^*) 
\Big( \cosh{(\Delta \Gamma/2) t}  -\sinh{(\Delta \Gamma/2) t}\Big) \Big] ~, \nl
(K_6(t) +\bar{K}_6(t))_{SM} &=& 0 ~.
\eea
We have ${\cal{A}}^{sh}_i = \mp {\cal{A}}^{ch}_i $
[Eq.~(\ref{effT1})], where the minus sign is for $i=1,2,5$, the plus
sign for $i=3$, and both quantities vanish when $i=4,6$. The effective
lifetimes are then predicted to be
\beq
\tau^{eff, i}_{B_s, SM} = \frac{\tau_{B_s}}{(1+y_s)} ~,~i=1,2,5 ~~,~~~~
\tau^{eff, i}_{B_s, SM} = \frac{\tau_{B_s}}{(1-y_s)} ~,~i=3 ~.
\label{efftSM}
\eeq
If the measurement of an effective lifetime differs from the SM
prediction, this will be a sign for NP \cite{fleischer}.

The SM untagged time-dependent angular distribution for $ \bspp$ takes
the form
\bea
\label{ADphiphiSM}
\frac{d^4 (\Gamma^{B_s} + \Gamma^{\bar{B}_s})}{dt d\vec{\omega} } 
&=& \frac{9}{32 \pi} \Big[{\cal{F}}_L (q^2,\vec{\omega}) {\cal{K}}_L(t)  + {\cal{F}}_H(q^2,\vec{\omega}) {\cal{K}}_H (t)\Big]~,
\eea
where the angular and time-dependent terms are
\bea
\label{GLGHSM}
{\cal{F}}_L (\vec{\omega}) &=&  \Big[|A_0|^2 f_1(\vec{\omega})+|A_\parallel|^2 f_2(\vec{\omega}) 
+ |A_0| |A_\parallel|\cos{(\delta_\parallel-\delta_0)}f_5(\vec{\omega})\Big]~,\nl
{\cal{F}}_H (\vec{\omega}) &=& |A_\perp|^2~,\nl
{\cal{K}}_L(t)  &=& 2 e^{-\Gamma_L t}~= 2e^{ - \Gamma t} 
\Big( \cosh{(\Delta \Gamma/2) t}  -\sinh{(\Delta \Gamma/2) t} \Big) ,\nl
{\cal{K}}_H(t)  &=& 2 e^{-\Gamma_H t}~=2 e^{ - \Gamma t} \Big( \cosh{(\Delta \Gamma/2) t}  +\sinh{(\Delta \Gamma/2) t} \Big) ~,
\eea
in which $(\delta_\parallel-\delta_0)= \arg(A_\| A_0^*)$.

Thus, if the $e^{-\Gamma_L t/2}$ and $e^{-\Gamma_H t/2}$ terms in the
time-dependent angular distribution [see Eq.~~(\ref{expLR})] can be
distinguished experimentally, the $|A_h|$ and
$\cos{(\delta_\parallel-\delta_0)}$ can be measured. However, as we
will see in the next subsection, these observables can be obtained
from time-integrated measurements.

\subsection{Untagged time-integrated distribution}

In the SM, the observables in the time-integrated untagged
distribution are
\bea
\label{UTKisTISM}
\langle K_1\rangle = \frac{\tau_{B_s}}{1+y_s} \, |A_0|^2 & ~~,~~~~ &
\langle K_2\rangle = \frac{\tau_{B_s}}{1+y_s} \, |A_\||^2 \nl
\langle K_3\rangle = \frac{\tau_{B_s}}{1-y_s} \, |A_\perp|^2 & ~~,~~~~ &
\langle K_4\rangle = 0 ~, \nl
\langle K_5\rangle = \frac{\tau_{B_s}}{1+y_s} \, |A_0| |A_\parallel|\cos{(\delta_\parallel-\delta_0)} & ~~,~~~~ &
\langle K_6\rangle = 0 ~.    
\eea
We have $y_s = 0.088 \pm 0.014$ and $\tau^{-1}_{B_s} = (0.6580 \pm
0.0085)~{\rm ps}^{-1}$ \cite{fleischer}. With this knowledge, the
$|A_h|$ and $\cos{(\delta_\parallel-\delta_0)}$ can be extracted from
the above measurements.  This is what CDF and LHCb have
presented \cite{CDFBs,LHCbphiphi}. 

\subsection{Polarization Fractions}

With no weak phases in the decay, we have $A_h = {\bar A}_h$, and the
$|A_h|^2$ can be measured in the untagged time-integrated distribution
[Eq.~(\ref{UTKisTISM})]. The polarization fractions are given by
\bea
f_0 &=& \frac{|A_0|^2}{|A_0|^2 +|A_\||^2 +|A_\perp|^2} ~,\nl
f_\| &=& \frac{|A_\||^2}{|A_0|^2 +|A_\||^2 +|A_\perp|^2} ~,\nl
f_\perp &=& \frac{|A_\perp|^2}{|A_0|^2 +|A_\||^2 +|A_\perp|^2} ~,
\eea
with total polarization $f_{tot} =f_0 + f_\| + f_\perp = 1$.

Now, in the presence of NP the distribution changes, and so the
experimental measurements have to be reinterpreted. We address this
issue in the next section.

\section{\boldmath $\bspp$ -- SM $+$ NP}

In this section, we consider NP contributions to $\bspp$, in the
mixing and/or in the decay. 

\subsection{ Polarization Fractions}

The polarization fractions can be written as
\bea
\label{PolarF}
f_0 &=& 
\frac{|A_0|^2 + |{\bar A}_0|^2}{|A_0|^2 + |{\bar A}_0|^2 +|A_\||^2 + |{\bar A}_\||^2
+|A_\perp|^2 + |{\bar A}_\perp|^2} = \frac{{\cal{A}}^{ch}_1}{\sum_{i =1,2,3}{\cal{A}}^{ch}_i } ~,\nl
f_\| &=& 
\frac{|A_\||^2 + |{\bar A}_\||^2}{|A_0|^2 + |{\bar A}_0|^2 +|A_\||^2 + |{\bar A}_\||^2
+|A_\perp|^2 + |{\bar A}_\perp|^2} = \frac{{\cal{A}}^{ch}_2}{\sum_{i =1,2,3}{\cal{A}}^{ch}_i } ~,\nl
f_\perp &=& 
\frac{|A_\perp|^2 + |{\bar A}_\perp|^2}{|A_0|^2 + |{\bar A}_0|^2 +|A_\||^2 + |{\bar A}_\||^2
+|A_\perp|^2 + |{\bar A}_\perp|^2} = \frac{{\cal{A}}^{ch}_3}{\sum_{i =1,2,3}{\cal{A}}^{ch}_i } ~.
\eea

In the above, the $f_h$ are written in terms of the $|A_h|^2$ and
$|{\bar A}_h|^2$. However, as noted above, what is measured
experimentally in the time-integrated untagged distribution are the
$\langle K_i\rangle$. It is therefore necessary to express the $f_h$
in terms of the $\langle K_i\rangle$. This is done as follows.  Using
Eq.~(\ref{effT3}), one can write
\bea
\label{AchshYs}
{\cal{A}}^{ch}_i &=& \frac{\langle K_i \rangle}{\tau_{B_s}}(1 + \eta_i y_s) Y_{i} ~,~~ i=1,2,3 ~,
\eea
where the quantity $Y_{i}$ is related to  $\tau^{eff, i}_{B_s}$ or ${\cal{A}}^{i}_{\Delta \Gamma}$:
\bea
\label{AchshYs2}
Y_i &=& \frac { 1} { (1+\eta_i y_s)} \Big(2 -\frac{\tau^{eff,
    i}_{B_s}}{\tau_{B_s}} (1 - y_s^2) \Big) = \frac{(1- \eta_i
  y_s)}{(1+ {\cal{A}}^{i}_{\Delta \Gamma} y_s)} ~,
\eea
with $\eta_{1,2} = 1$, and $\eta_{3} = -1$.  
{}From Eq.~(\ref{UTKisTI}) we have ${\cal{A}}^{i}_{\Delta \Gamma} = {{\cal{A}}^{sh}_i}/{{\cal{A}}^{ch}_i} = $
\bea
\label{AiDGcases}
\begin{cases}
{-2\Big({\rm Re}(A_0^* \bar{A}_0) \cos \phi_s +~{\rm
    Im}(A_0^* \bar{A}_0) \sin \phi_s \Big)}/{\Big( |A_0|^2 + |{\bar
    A}_0|^2 \Big)} ~~,~~ i = 1 ~, \\
{-2\Big({\rm Re}(A_\|^* \bar{A}_\|) \cos \phi_s +~{\rm
    Im}(A_\|^* \bar{A}_\|) \sin \phi_s \Big)}/{\Big( |A_\||^2 + |{\bar
    A}_\||^2 \Big)} ~~,~~ i = 2 ~, \\
{2\Big({\rm Re}(A_\perp^* \bar{A}_\perp) \cos \phi_s +~{\rm
    Im}(A_\perp^* \bar{A}_\perp) \sin \phi_s \Big)}/{\Big( |A_\perp|^2 + |{\bar
    A}_\perp|^2 \Big)} ~~,~~ i = 3 ~.
\end{cases}
\eea

In the SM, the weak phases of the $A_h$ vanish and $\phi_s = 0$, so
that ${\cal{A}}^{i}_{\Delta \Gamma} = \pm 1$ (the minus sign is for
$i=1,2$, and the plus sign is for $i=3$). This implies that
$Y_{1,2,3}=1$, so that the polarization fractions are
\bea
\label{PolarFSM}
f_0^{SM} &=& \frac{\langle K_1 \rangle (1+y_s) }{\langle K_1 \rangle
  (1+y_s) + \langle K_2 \rangle (1+y_s) + \langle K_3 \rangle ( 1-y_s)} ~,\nl
f_\|^{SM} &=& \frac{\langle K_2 \rangle (1+y_s) }{\langle K_1 \rangle
  (1+y_s) + \langle K_2 \rangle (1+y_s) + \langle K_3 \rangle ( 1-y_s)} ~,\nl
f_\perp^{SM} &=&\frac{\langle K_3 \rangle (1-y_s) }{\langle K_1 \rangle
  (1+y_s) + \langle K_2 \rangle (1+y_s) + \langle K_3 \rangle ( 1-y_s)} ~.
\eea
Note that these are consistent with Eq.~(\ref{UTKisTISM}).
However, if there is NP in the mixing and/or the decay, we have
$Y_{1,2,3} \ne 1$, so that the polarization fractions take the form
\bea
\label{PolarFSMplusNP}
f_0 &=& \frac{\langle K_1 \rangle (1+y_s)Y_1 }{\langle K_1 \rangle (1+y_s)Y_1 + \langle K_2 \rangle (1+y_s)Y_2 + \langle K_3 \rangle ( 1-y_s)Y_3  }  ~,\nl
f_\| &=& \frac{\langle K_2 \rangle (1+y_s)Y_2 }{\langle K_1 \rangle (1+y_s)Y_1 + \langle K_2 \rangle (1+y_s)Y_2 + \langle K_3 \rangle ( 1-y_s)Y_3  }  ~,\nl
f_\perp &=&\frac{\langle K_3 \rangle (1-y_s)Y_3 }{\langle K_1 \rangle (1+y_s)Y_1 + \langle K_2 \rangle (1+y_s)Y_2 + \langle K_3 \rangle ( 1-y_s)Y_3  }  ~.
\eea
The $f_h$ are expressed completely in terms of measured quantities.
The $\langle K_i \rangle$'s are obtained from the untagged angular
distribution, and one can calculate the $Y_i$ using the measured
effective lifetimes. If the effective lifetimes have not been measured
then, ${\cal{A}}^{i}_{\Delta \Gamma}$ can be varied within a certain
range to get a range for the $Y_i$.

Thus, to obtain the correct polarization fractions in the presence of
NP, Eq.~(\ref{PolarFSMplusNP}), which includes factors of $Y_i$, must
be used. This is one of the main points of the paper.  However,
experiments have used Eq.~(\ref{PolarFSM}), so they have effectively
excluded NP. If this possibility is allowed, the analysis must be
redone and we discuss this in Sec.~5.

The difference between Eqs.~(\ref{PolarFSM}) and
(\ref{PolarFSMplusNP}) is related to the difference $Y_i - 1$.  One
can see from Eq.~(\ref{AchshYs2}) that $Y_i - 1 \to 0$ in the limit
that $y_s \to 0$. This indicates that $f_h - f_h^{SM} = O(y_s)$. Since
$y_s = 0.088 \pm 0.014$, this corresponds to a correction to the
polarization fractions of $O(10\%)$. This is not large, but it may be
important given that the measurements hope to identify the presence of
NP.

\subsection{Other Observables}

In Sec.~\ref{VVangdist}, we noted that the angular distribution of the
decay $B^0_q \to V_1 V_2$ ($q=d,s$) is proportional to $\sum^6_{i=1}
K_i(t) f_i(\vec{\omega})$, where $\vec{\omega} = (\cos{\theta_1
},\cos{\theta_2 }, \Phi)$ [Eq.~(\ref{AD:Bsphiphi})]. In the previous
subsection, we discussed polarization fractions, observables which are
dependent on $\langle K_i \rangle$, $i=1,2,3$. We now turn to $i=4,6$.

In the present case, $K_4$ and $K_6$ are related to the triple
products (TP's) in $\bspp$. The expressions for the untagged
observables in $B^0_{d,s} \to V_1 V_2$ are given in
Eq.~(\ref{UTKis}). For convenience, $K_i(t)+\bar{K}_i(t)$ ($i=4,6$)
are repeated below:
\bea
K_4(t) +\bar{K}_4(t) &=&  e^{ - \Gamma_s t}  \Big[ \Big( {\rm Im} ( A_\perp A_\|^* ) - {\rm Im} ( \bar{A}_\perp \bar{A}_\|^* ) \Big) \cosh{(\Delta \Gamma_s/2) t} \nl 
&&  -~ \Big(( {\rm Im} ( A_\perp \bar{A}_\|^* ) - {\rm Im} ( \bar{A}_\perp A_\|^* ) )\cos \phi_s \nl 
&& +~( {\rm Re} ( A_\perp \bar{A}_\|^* ) + {\rm Re} ( \bar{A}_\perp A_\|^* ) ) \sin \phi_s\Big)
\sinh{(\Delta \Gamma_s/2) t}  \Big] ~, \nl
K_6(t) +\bar{K}_6(t) &=&  e^{ - \Gamma_s t}  \Big[ \Big( {\rm Im} ( A_\perp A_0^* ) - {\rm Im} ( \bar{A}_\perp \bar{A}_0^* ) \Big) 
\cosh{(\Delta \Gamma_s/2) t} \nl 
&&  -~ \Big(( {\rm Im} ( A_\perp \bar{A}_0^* ) - {\rm Im} ( \bar{A}_\perp A_0^* ) )\cos \phi_s \nl 
&& +~( {\rm Re} ( A_\perp \bar{A}_0^* ) + {\rm Re} ( \bar{A}_\perp A_0^* ) ) \sin \phi_s\Big)
\sinh{(\Delta \Gamma_s/2) t}  \Big] ~.
\eea
Now, as discussed earlier, in the SM the weak phases in $\bspp$, both
in the mixing and in the decay, are all approximately zero, so that
$K_4(t)+\bar{K}_4(t)$ and $K_6(t)+\bar{K}_6(t)$ vanish. Thus, if one
finds evidence for a nonzero TP, this is a clear sign of NP.

Suppose first that there is NP, with a nonzero weak phase, only in the
mixing. In this case, the first two terms in each of
$K_i(t)+\bar{K}_i(t)$ ($i=4,6$) are zero, but the third is
nonzero. This is a particularly interesting situation, as it
corresponds to a TP generated through mixing. It arises only because
$\Delta \Gamma_s$ is nonzero; mixing-induced TP's cannot be produced
in $\bd$ decays. And, although $\Delta \Gamma_s\ne 0$, it is still
not large, so that the associated TP is also rather small.

The second possibility is that there is NP, with a nonzero weak phase,
only in the decay. In this case, the first two terms in each of
$K_i(t)+\bar{K}_i(t)$ ($i=4,6$), proportional to $\cosh{(\Delta
  \Gamma_s/2) t}$ and $\cos \phi_s = 1$, are nonzero, but the third is
zero. And of course one can have NP in both the mixing and the
decay. If a TP is seen, its source can be determined through its time
dependence.

Both $K_i(t)+\bar{K}_i(t)$ ($i=4,6$) are CP-violating, so they
correspond to true TP's. They can be nonzero only if there are two
interfering amplitudes with a relative weak phase. If there is NP in
the mixing, the amplitudes are $A(\bspp)$ and
$A(\bs\to\bsbar\to\phi\phi)$; if there is NP in the decay, the
amplitudes are $A(\bspp)_{SM}$ and $A(\bspp)_{NP}$. In addition, in
order to produce a TP, the two interfering amplitudes must be
kinematically different \cite{TPDL}. For the case of NP in the decay,
this is satisfied straightforwardly. But for NP in the mixing, how are
$\bspp$ and $\bs\to\bsbar\to\phi\phi$ kinematically different? The
point is that mixing-induced TP's are generated due to a nonzero
$\Delta \Gamma_s$. That is, although $\bspp$ is a penguin decay,
$\bsbar\to\phi\phi$ occurs via a mechanism which contributes to
$\Delta \Gamma_s$. For example, one possibility is the $\bs \to
\bsbar$ transition via the intermediate states $D_s^{*+}D_s^{*-}$
\cite{ALOPR}, with the $\bsbar$ decaying to $\phi\phi$. The $\bs$ and
$\bsbar$ decays are clearly kinematically different.

We now turn to the measurement of TP's. Here we focus on the
time-integrated untagged observables, $\langle K_i\rangle$. We have
$\langle K_i \rangle \propto {\cal{A}}^{ch}_i + {\cal{A}}^{sh}_i y_s$
[Eq.~(\ref{effT1})]. Specifically, the $\langle K_{4,6} \rangle$ are
given in Eq.~(\ref{UTKisTI}):
\bea
\label{theoTP}
\langle K_4\rangle &=&  \frac{\tau_{B_s}}{2(1-y^2_s)}   \Big[ \Big( {\rm Im} ( A_\perp A_\|^* ) - {\rm Im} ( \bar{A}_\perp \bar{A}_\|^* ) \Big) -~ \Big(( {\rm Im} ( A_\perp \bar{A}_\|^* ) - {\rm Im} ( \bar{A}_\perp A_\|^* ) )\cos \phi_s \nl 
&& \hskip2.5truecm  
+~( {\rm Re} ( A_\perp \bar{A}_\|^* ) + {\rm Re} ( \bar{A}_\perp A_\|^* ) ) \sin \phi_s\Big)
y_s \Big] ~, \nl
\langle K_6\rangle &=& \frac{\tau_{B_s}}{2(1-y^2_s)} \Big[ \Big( {\rm Im} ( A_\perp A_0^* ) - {\rm Im} ( \bar{A}_\perp \bar{A}_0^* ) \Big) -~ \Big(( {\rm Im} ( A_\perp \bar{A}_0^* ) - {\rm Im} ( \bar{A}_\perp A_0^* ) )\cos \phi_s \nl 
&& \hskip2.5truecm 
+~( {\rm Re} ( A_\perp \bar{A}_0^* ) + {\rm Re} ( \bar{A}_\perp A_0^* ) ) \sin \phi_s\Big)
y_s  \Big] ~.
\eea 
The TP's in the untagged distribution can be measured by constructing
asymmetries involving the angular variables. We start by integrating
Eq.~(\ref{ADphiphi-inte}) over $\theta_1$ and $\theta_2$ to obtain
the differential rate:
\bea
\label{ADphiphi-inte2}
\frac{d \langle\Gamma(B^0_q \to V_1 V_2)\rangle}{ d\Phi }  &=& \frac{1}{2 \pi} \Big[\langle K_1\rangle   +  2 \langle K_2\rangle \cos^2{\Phi}  + 2\langle K_3\rangle \sin^2{\Phi} - 2 \langle K_4\rangle\sin{2 \Phi} \Big]~.
\eea
Note that the time-integrated untagged decay rate can be obtained by
integrating out the azimuthal angle $\Phi$:
\bea
\label{Undecayrate}
\langle\Gamma(B^0_q \to V_1 V_2)\rangle &=& \Big[\langle K_1\rangle   +  \langle K_2\rangle  + \langle K_3\rangle  \Big]~.
\eea


Following Ref.~\cite{GRTP} we can define asymmetries to measure the
TP's.  We begin with $i=4$, for which $f_4(\vec{\omega}) =
-2\sin^2{\theta_1}\sin^2{\theta_2} \sin{2\Phi}$. We define $u \equiv
\sin{2 \Phi}$. The TP asymmetry between the number of decays involving
positive and negative values of $u$ is given by \cite{GRTP,TPDL}
\bea
\label{Audef}
{\cal{A}}_u &=& \frac{1}{2} \Big[\frac{ \langle\Gamma(\bspp), u >0
    \rangle - \langle\Gamma(\bspp), u<0 \rangle}{\langle\Gamma(\bspp),
    u >0 \rangle + \langle\Gamma(\bspp), u <0 \rangle} \Big] \nl
&=&  -\frac{2}{\pi} [{\cal{A}}^{(2)}_T]_{exp} ~~,~~~~
[{\cal{A}}^{(2)}_T]_{exp} = \frac{\langle K_4 \rangle}{\langle\Gamma(\bspp)\rangle  }~.
\eea
As noted above, if ${\cal{A}}_u \ne 0$ is found, this would clearly
indicate NP.  However, we would like to know the relation between
$[{\cal{A}}^{(2)}_T]_{exp}$ and the theoretical expression for the TP
in Eq.~(\ref{theoTP}). The measured TP $[{\cal{A}}^{(2)}_T]_{exp}$ is
related to $[{\cal{A}}^{(2)}_T]_{theo}$ via
\bea
\label{AT2rel1}
[{\cal{A}}^{(2)}_T]_{exp}&=& [{\cal{A}}^{(2)}_T]_{theo}\frac{\tau_{B_s}}{\langle\Gamma(\bspp)\rangle} \frac{(1 + A^{(4)}_{\Delta \Gamma} y_s)}{ (1 - y_s^2)} ~,
\eea
where $A^{(4)}_{\Delta \Gamma} =
{\cal{A}}^{sh}_{4}/{\cal{A}}^{ch}_{4}$ and
\bea
\label{AT2theo1}
[{\cal{A}}^{(2)}_T]_{theo} &=& {\cal{A}}^{ch}_4 = \frac12 \left( {\rm
  Im} ( A_\perp A_\|^* ) - {\rm Im} ( \bar{A}_\perp \bar{A}_\|^*)
\right) ~.
\eea
If we define the dimensionless theoretical TP as
\bea
{\cal{TP}}_2 & \equiv & [{\cal{A}}^{(2)}_T]_{theo}\frac{\tau_{B_s}}{\langle\Gamma(\bspp)\rangle} ~,
\label{dimTP2}
\eea
Eq.~(\ref{AT2rel1}) details the corrections to the naive relation
$[{\cal{A}}^{(2)}_T]_{exp} = {\cal{TP}}_2$ due to a nonzero (NP)
${\cal{A}}^{sh}_{4}$. (In the SM, $[{\cal{A}}^{(2)}_T]_{theo} = 0$, so
the relation is trivial.)

For $i=6$ we have $f_6(\vec{\omega}) = - \sqrt{2}\sin{2\theta_1}
\sin{2\theta_2} \sin{\Phi}$. We define \\ $v \equiv
{\rm sign}(\cos{\theta_1}\cos{\theta_2})\sin{\Phi}$, which has the following
associated TP asymmetry \cite{GRTP}:
\bea
\label{Avdef}
{\cal{A}}_v &=& \frac{1}{2} \Big[\frac{ \langle\Gamma(\bspp), v >0 \rangle - \langle\Gamma(\bspp), v<0 \rangle}{\langle\Gamma(\bspp), v >0 \rangle + \langle\Gamma(\bspp), v <0 \rangle} \Big] \nl 
&=&  -\frac{\sqrt{2}}{\pi} [{\cal{A}}^{(1)}_T]_{exp} ~~,~~~~
[{\cal{A}}^{(1)}_T]_{exp} = \frac{\langle K_6 \rangle}{\langle\Gamma(\bspp)\rangle  }~.
\eea
Then
\bea
\label{AT1rel1}
[{\cal{A}}^{(1)}_T]_{exp}&=& [{\cal{A}}^{(1)}_T]_{theo}\frac{\tau_{B_s}}{\langle\Gamma(\bspp)\rangle} \frac{(1 + A^{(6)}_{\Delta \Gamma} y_s)}{ (1 - y_s^2)} ~,
\eea
where $A^{(6)}_{\Delta \Gamma} =
{\cal{A}}^{sh}_{6}/{\cal{A}}^{ch}_{6}$ and
\bea
\label{AT1theo1}
[{\cal{A}}^{(1)}_T]_{theo} &=& {\cal{A}}^{ch}_6 = \frac12 \left( {\rm
  Im} ( A_\perp A_0^* ) - {\rm Im} ( \bar{A}_\perp \bar{A}_0^*)
\right) ~.
\eea
We can again define the dimensionless theoretical TP as
\bea
{\cal{TP}}_1 & \equiv & [{\cal{A}}^{(1)}_T]_{theo}\frac{\tau_{B_s}}{\langle\Gamma(\bspp)\rangle} ~.
\label{dimTP1}
\eea
Eq.~(\ref{AT1rel1}) gives the corrections to the
naive relation $[{\cal{A}}^{(1)}_T]_{exp} = {\cal{TP}}_1$.

Finally, we turn to $i=5$, which corresponds to a CP-conserving
observable. From Eq.~(\ref{UTKisTI}),
\bea
\langle K_5\rangle &=& \frac{\tau_{B_s}}{2(1-y^2_s)}  \Big[ \Big( {\rm Re}(A_\| A_0^*) + {\rm Re}(\bar{A}_\| \bar{A}_0^*) \Big) 
          -~\Big( ( {\rm Re}(A_\| \bar{A}_0^*) + {\rm Re}(\bar{A}_\| A_0^*) ) 
               \cos \phi_s  \nl 
&& \hskip2.5truecm 
         -~( {\rm Im}(A_\| \bar{A}_0^*) - {\rm Im}(\bar{A}_\| A_0^*) ) 
              \sin \phi_s  \Big) y_s \Big] ~.
\eea 
We have $f_5(\vec{\omega}) =\sqrt{2}\sin{2\theta_1} \sin{2\theta_2}
\cos{\Phi}$, so we define $w \equiv
sign(\cos{\theta_1}\cos{\theta_2})\cos{\Phi}$ The associated asymmetry
is
\bea
\label{Awdef}
{\cal{A}}_w &=& \frac{1}{2} \Big[\frac{ \langle\Gamma(\bspp), w >0 \rangle - \langle\Gamma(\bspp), w<0 \rangle}{\langle\Gamma(\bspp ), w >0\rangle + \langle\Gamma(\bspp), w <0 \rangle} \Big] \nl 
&=& \frac{\sqrt{2}}{\pi} [A^{(5)}]_{exp} ~~,~~~~ 
[A^{(5)}]_{exp} = {\langle K_5 \rangle}/{\langle\Gamma(\bspp)\rangle } ~.
\eea
We have
\bea
\label{A5rel}
[A^{(5)}]_{exp} &=& [A^{(5)}]_{theo} \,
\frac{\tau_{B_s}}{\langle\Gamma(\bspp)\rangle} \frac{(1 +
  A^{(5)}_{\Delta \Gamma} y_s)}{ (1 - y_s^2)} \nl
&=& [A^{(5)}]_{theo}\frac{\tau_{B_s}}{\langle\Gamma(\bspp)\rangle} \Big(2 -\frac{\tau^{eff,5}_{B_s}}{\tau_{B_s}}  (1 - y_s^2) \Big) ~,
\eea
where $A^{(5)}_{\Delta \Gamma} =
{\cal{A}}^{sh}_{5}/{\cal{A}}^{ch}_{5}$, the effective lifetime
$\tau^{eff,5}_{B_s}$ is defined in Eq.~(\ref{effT2}), and
\bea
\label{ARthe}
[A^{(5)}]_{theo} & = & {\cal{A}}^{ch}_5 = \frac{1}{2}   \Big( {\rm Re}(A_\| A_0^*) + {\rm Re}(\bar{A}_\| \bar{A}_0^*) \Big)  ~.
\eea  

\subsection{NP Parameters}

12 observables can be measured from the time-dependent untagged
angular distribution (Sec.~\ref{untagged}). With these, one can
identify if NP is present in the mixing and/or the decay.  However, we
will also want to identify its properties. To be specific, if there is
NP in the decay amplitude, it will be important to measure the various
NP parameters. With this in mind, the question is: how many
theoretical unknowns are there in the most general SM $+$ NP $\bspp$
amplitude? If there are fewer than 12, then we can extract all the
unknowns.

In writing the SM $+$ NP $\bspp$ amplitude, we have the following
points:
\begin{itemize}

\item The SM weak phases are $\simeq 0$.

\item Assuming that the NP amplitudes satisfy $|{\cal{A}}^{NP}_h|
  <|{\cal{A}}^{SM}_h|$, the NP strong phases are negligible
  \cite{dattaNP1}. This means that if there are many NP amplitudes,
  they can all be combined into a single term with an effective
  magnitude and weak phase.

\item In the heavy-quark limit, we have
  ${\cal{A}}^{SM}_{\perp}=-{\cal{A}}^{SM}_{\|}$ \cite{KaganBRY}.

\end{itemize}
Taking these points into account, the most general SM $+$ NP $\bspp$
helicity amplitude can then be written
\beq
\label{BVVAmpsNewdef2}
{\cal{A}}_h = |{\cal{A}}^{SM}_h| e^{i \delta^{SM}_h} +  |{\cal{A}}^{NP}_h| e^{i \phi_h} ~.
\eeq
There are a total of 11 unknown theoretical parameters -- 5 magnitudes
(2 SM, 3 NP), 2 SM strong phases, 3 NP weak phases, and the
mixing phase $\phi_s$. In principle, these can all be extracted from
the 12 observables.

However, note that Eq.~(\ref{BVVAmpsNewdef2}) includes a different NP weak
phase $\phi_{h}$ for each helicity amplitude. But in many NP models
the weak phases are helicity independent. In this case there is only
one NP weak phase $\phi$, and the number of theoretical unknowns is
reduced to 9. This is a model-dependent result, but it is still very
general.

Finally, if the NP is purely left-handed or right-handed, then
${\cal{A}}^{NP}_{\perp}=\mp {\cal{A}}^{NP}_{\|}$ \cite{fakeTP}, which
further reduces the number of theoretical unknowns by one.

In all cases, assuming the time-dependent untagged angular
distribution can be measured, there are more observables than
unknowns, and so we will be able to extract all the NP parameters in
the decay. In this way, we may be able to identify the type of NP that
is present.

\section{Numerical Analysis}

Recently, the CDF and LHCb Collaborations have reported measurements
for the polarization amplitudes, the strong-phase difference between
$A_\|$ and $A_0$, and the triple-product asymmetries in $\bspp$. The
LHCb results \cite{LHCbphiphi} are summarized in Table~\ref{LHCres}.
The values are in good agreement with those reported by the CDF
Collaboration \cite{CDFBs}, except for the TP's, though all
measurements are consistent within errors.

\begin{table}[tbh]
\center
\begin{tabular}{cc}
\hline
\hline
Observable & Measurement \\ \hline
$|A_0|_{exp}^2$  & $ 0.365 \pm 0.022~({\rm stat}) \pm 0.012~({\rm syst})$ \\
$|A_\perp|_{exp}^2$  & $ 0.291 \pm 0.024~({\rm stat}) \pm 0.010~({\rm syst})$ \\
$|A_\||_{exp}^2$ & $ 0.344 \pm 0.024~({\rm stat}) \pm 0.014~({\rm syst})$ \\
$\cos(\delta_\parallel - \delta_0)$ & $ -0.844 \pm 0.068~({\rm stat}) \pm 0.029~({\rm syst})$ \\
${\cal{A}}_u$ & $ -0.055 \pm 0.036~({\rm stat}) \pm 0.018~({\rm syst})$ \\
${\cal{A}}_v$ & $ 0.010 \pm 0.036~({\rm stat}) \pm 0.018~({\rm syst})$ \\
\hline
\hline
\end{tabular}
\caption{Measured polarization amplitudes, strong-phase difference,
  and triple-product asymmetries in $\bspp$ \cite{LHCbphiphi}. The sum
  of the $|A_h|_{exp}^2$ terms is constrained to unity.}
\label{LHCres}
\end{table}

The experiments have measured the $\langle K_i\rangle$ and constructed
the polarization fractions assuming the SM. As discussed previously,
if one allows for the possibility of NP in $\btos$ transitions, this
analysis must be modified. This is done below.

We denote the measured value of $y_s$ as $y_{s0}$.  From
Eq.~(\ref{UTKisTISM}) we have
\bea
\langle K_1\rangle &=& \frac{\tau_{B_s}}{1+y_s} \, |A_0|^2 
=\frac{\tau_{B_s}}{1+y_{s0}} \, |A_0|_{y_s=y_{s0}}^2 ~, \nl 
\langle K_2\rangle &=& \frac{\tau_{B_s}}{1+y_s} \, |A_\||^2 
=\frac{\tau_{B_s}}{1+y_{s0}} \, |A_\||_{y_s=y_{s0}}^2 ~, \nl 
\langle K_3\rangle &=& \frac{\tau_{B_s}}{1-y_s} \, |A_\perp|^2 
=\frac{\tau_{B_s}}{1-y_{s0}} \, |A_\perp|_{y_s=y_{s0}}^2 ~.
\label{asquared}
\eea
The experimental measurements in Table~\ref{LHCres} are then
\bea
|A_0|_{exp}^2 & = & \frac{|A_0|_{y_s=y_{s0}}^2}
{ |A_0|_{y_s=y_{s0}}^2+ |A_\||_{y_s=y_{s0}}^2 +|A_\perp|_{y_s=y_{s0}}^2} ~, \nl
 |A_\||_{exp}^2 & = & \frac{|A_\||_{y_s=y_{s0}}^2}
{ |A_0|_{y_s=y_{s0}}^2+|A_\||_{y_s=y_{s0}}^2+|A_\perp|_{y_s=y_{s0}}^2} ~, \nl
|A_\perp|_{exp}^2 & = & \frac{|A_\perp|_{y_s=y_{s0}}^2}
{ |A_0|_{y_s=y_{s0}}^2+ |A_\||_{y_s=y_{s0}}^2 +|A_\perp|_{y_s=y_{s0}}^2} ~.
\label{measurement}
\eea
One can now calculate the polarization fractions in the SM as a
function of $y_s$.  Inputting the expressions for the $\langle
K_i\rangle $ from Eq.~(\ref{asquared}) into Eq.~(\ref{PolarFSM}), and
using Eq.~(\ref{measurement}), we obtain
\bea
f_0^{SM} & = & \frac{|A_0|_{exp}^2 \frac{1+y_s}{1+y_{s0}}}
{|A_0|_{exp}^2 \frac{1+y_s}{1+y_{s0}}+|A_\||_{exp}^2 \frac{1+y_s}{1+y_{s0}}
+|A_\perp|_{exp}^2 \frac{1-y_s}{1-y_{s0}}} ~, \nl
f_{\|}^{SM} & = & \frac{|A_\||_{exp}^2 \frac{1+y_s}{1+y_{s0}}}
{|A_0|_{exp}^2 \frac{1+y_s}{1+y_{s0}}+|A_\||_{exp}^2 \frac{1+y_s}{1+y_{s0}}
+|A_\perp|_{exp}^2 \frac{1-y_s}{1-y_{s0}}} ~, \nl
f_{\perp}^{SM} & = & \frac{|A_\perp|_{exp}^2 \frac{1-y_s}{1-y_{s0}}}
{|A_0|_{exp}^2 \frac{1+y_s}{1+y_{s0}}+|A_\||_{exp}^2 \frac{1+y_s}{1+y_{s0}}
+|A_\perp|_{exp}^2 \frac{1-y_s}{1-y_{s0}}} ~.
\label{SMpolarFversusys}
\eea
Hence the $|A_i|_{exp}^2$ in Table~\ref{LHCres} are just the
$f_i^{SM}$ defined in Eq.~(\ref{SMpolarFversusys}) with $y_s =y_{s0}$.

\begin{figure}[h]
\centering
\subfigure{
\includegraphics[height=3.6cm]{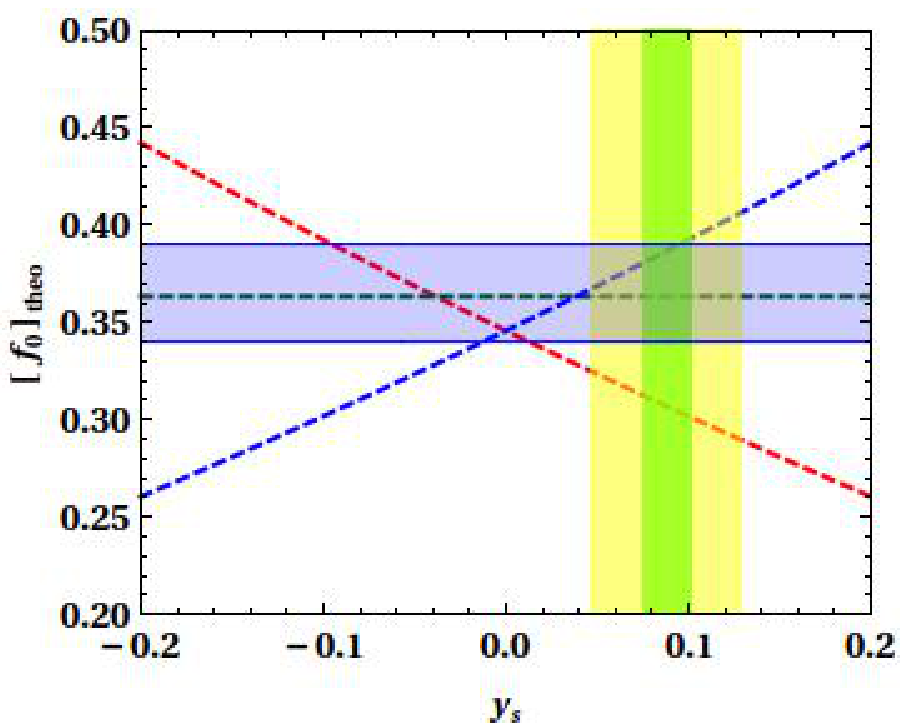}
}
\subfigure{
\includegraphics[height=3.6cm]{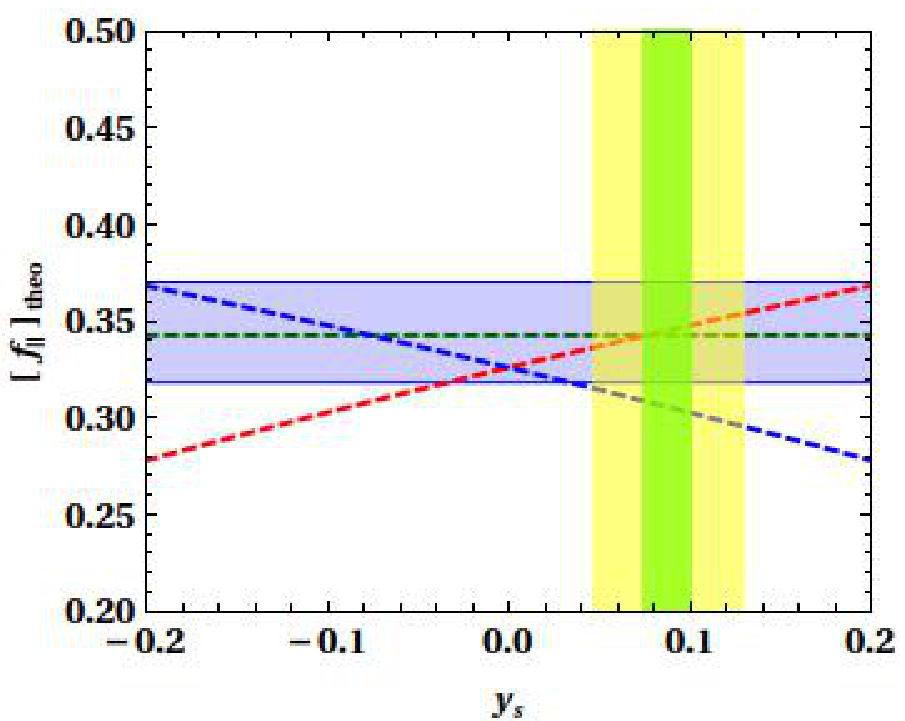}
}
\subfigure{
\includegraphics[height=3.6cm]{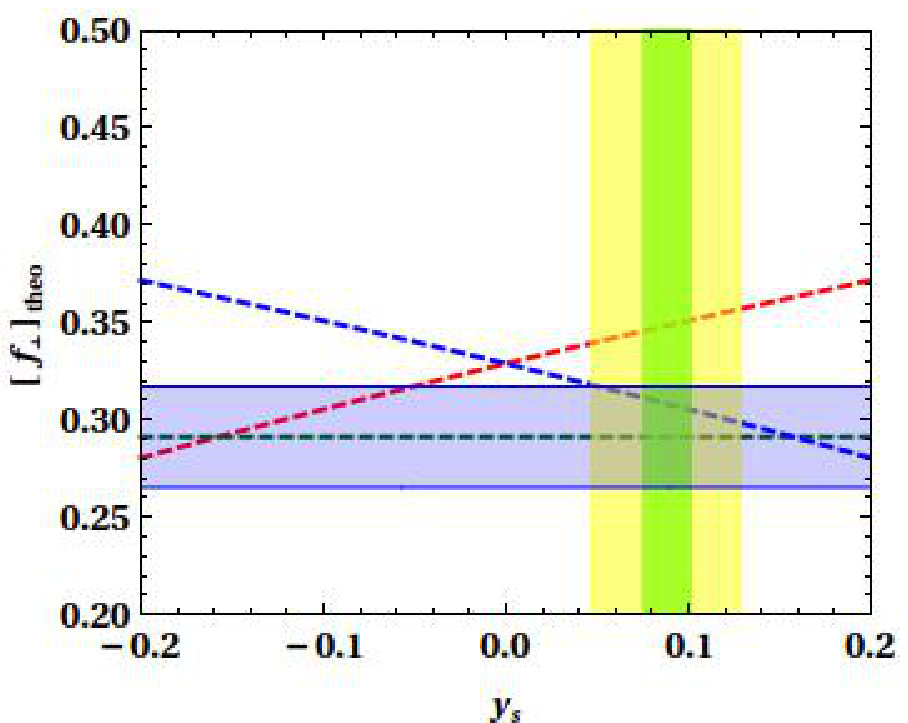}
}
\caption{The dependence of the theoretical polarization fractions
  $f_0$, $f_\parallel$ and $f_\perp$ on the decay width parameter
  $y_s$ for different values of ${\cal{A}}^{1,2,3}_{\Delta
    \Gamma}$. The red line corresponds to $(A^{1}_{\Delta \Gamma} = 1,
  A^{2}_{\Delta \Gamma} =-1, A^{3}_{\Delta \Gamma} =-1)$, while the
  blue line has $(A^{1}_{\Delta \Gamma} =-1, A^{2}_{\Delta \Gamma} =1,
  A^{2}_{\Delta \Gamma} =1)$.  In all plots the experimental result
  $[|A|^2_{h =0, \parallel,\perp}]_{exp}$ (horizontal region) is
  allowed to vary by $\pm 1\sigma$ (see Table~\ref{LHCres}).  The
  vertical bands correspond to $y_s$, with $\pm 1 \sigma$ (green) or
  $\pm 3 \sigma$ (yellow) errors.}
  \label{fig:pol}
\end{figure}

The true polarization fractions can then be obtained by inputting the
expressions for the $\langle K_i\rangle $ from Eq.~(\ref{asquared})
into Eq.~(\ref{PolarFSMplusNP}), and using Eq.~(\ref{measurement}):
\bea
f_0 & = & \frac{|A_0|_{exp}^2 \frac{1+y_s}{1+y_{s0}}Y_1}
{|A_0|_{exp}^2 \frac{1+y_s}{1+y_{s0}}Y_1+|A_\||_{exp}^2 \frac{1+y_s}{1+y_{s0}}Y_2
+|A_\perp|_{exp}^2 \frac{1-y_s}{1-y_{s0}}Y_3} ~, \nl
f_{\|} & = & \frac{|A_\||_{exp}^2 \frac{1+y_s}{1+y_{s0}}Y_2}
{|A_0|_{exp}^2 \frac{1+y_s}{1+y_{s0}}Y_1+|A_\||_{exp}^2 \frac{1+y_s}{1+y_{s0}}Y_2
+|A_\perp|_{exp}^2 \frac{1-y_s}{1-y_{s0}}Y_3} ~, \nl
f_{\perp} & = & \frac{|A_\perp|_{exp}^2 \frac{1-y_s}{1-y_{s0}}Y_3}
{|A_0|_{exp}^2 \frac{1+y_s}{1+y_{s0}}Y_1+|A_\||_{exp}^2 \frac{1+y_s}{1+y_{s0}}Y_2
+|A_\perp|_{exp}^2 \frac{1-y_s}{1-y_{s0}}Y_3} ~.
\label{polarFversusys}
\eea
In Fig.~\ref{fig:pol} we plot the dependence of the polarization
fractions $f_0$, $f_\parallel$ and $f_\perp$ as a function of
$y_s$. This figure is read as follows. In all plots the horizontal
region represents the experimental result, in which $[|A|^2_{h =0,
    \parallel,\perp}]_{exp}$ is allowed to vary by $\pm 1\sigma$ (see
Table~\ref{LHCres}).  Also, the vertical bands correspond to $y_s$,
with $\pm 1 \sigma$ (green) or $\pm 3 \sigma$ (yellow) errors. In the
SM we have $Y_i=1$, corresponding to $(A^{1}_{\Delta \Gamma} = -1,
A^{2}_{\Delta \Gamma} =-1, A^{3}_{\Delta \Gamma} =1)$
[Eq.~(\ref{AiDGcases})]. In order to illustrate the effect of NP, we
take $(A^{1}_{\Delta \Gamma} = 1, A^{2}_{\Delta \Gamma} =-1,
A^{3}_{\Delta \Gamma} =-1)$ (red line) or $(A^{1}_{\Delta \Gamma} =-1,
A^{2}_{\Delta \Gamma} =1, A^{2}_{\Delta \Gamma} =1)$ (blue line). For
these values of $A^{i}_{\Delta \Gamma}$, we have $Y_i\ne 1$. Consider
first $f_0$. In the SM the experimental measurement implies $0.33 \le
f_0 \le 0.40$. However, with NP, the value of $f_0$ can lie outside
this range -- for example, on the red line it can be as small as
0.29. The behavior is similar for $f_\parallel$ and $f_\perp$. This
shows explicitly that, in the presence of NP, the $\bspp$ polarization
fractions can be changed from their SM values by $O(10\%)$ for the
current value of $y_s$.

\begin{figure}[t]
\begin{center}
\includegraphics[width=6cm]{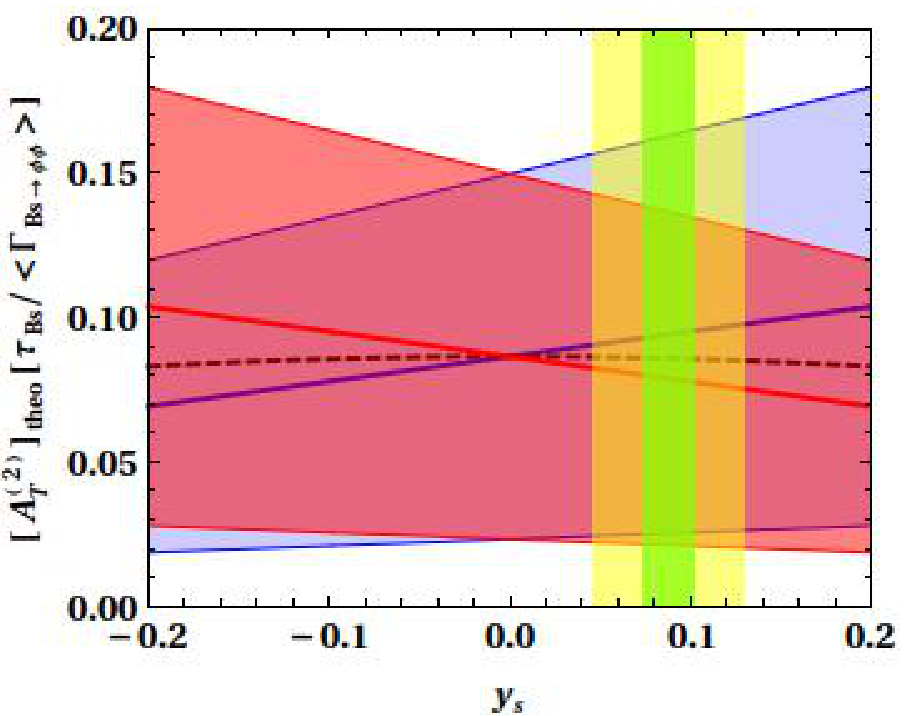}~~~~~~~\includegraphics[width=6cm]{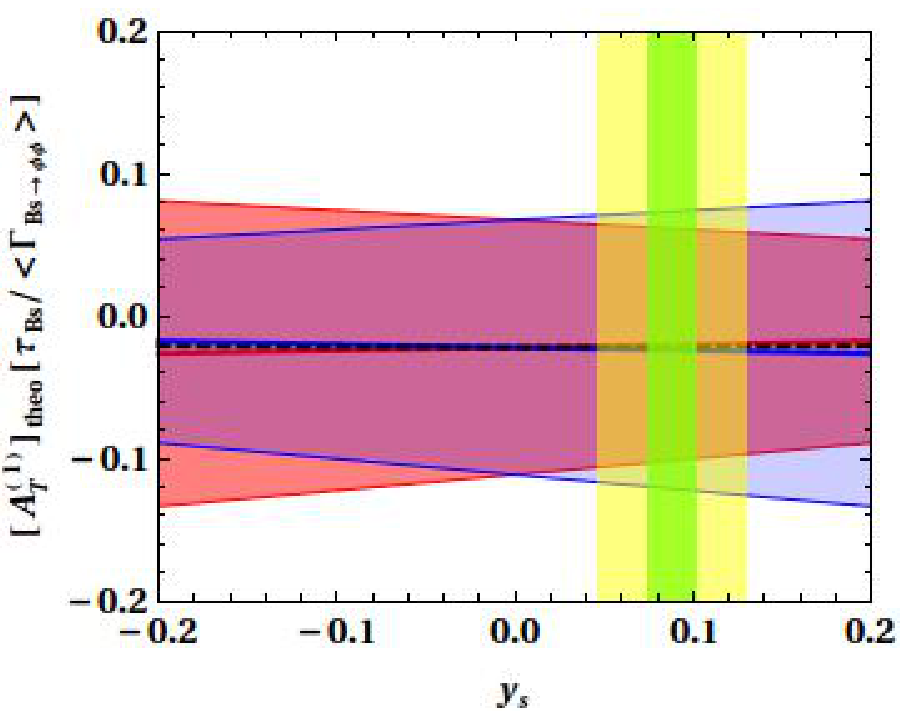}
\end{center}
\caption{The dependence of the theoretical TP's
  $[A^{(2)}_T]_{theo}{\tau_{B_s}}/{\langle\Gamma(\bspp)\rangle}$
  (left) and
  $[A^{(1)}_T]_{theo}{\tau_{B_s}}/{\langle\Gamma(\bspp)\rangle}$
  (right) on $y_s$ for different values of $A^{4(6)}_{\Delta \Gamma}$.
  In the red and blue regions, we take $A^{4(6)}_{\Delta \Gamma} = \pm
  1$, respectively. Also, ${\cal{A}}_u$ (left) and ${\cal{A}}_v$
  (right) are allowed to vary by $\pm 1\sigma$ (see
  Table~\ref{LHCres}). The dashed black lines correspond to the
  central values of ${\cal{A}}_u$ (left) and ${\cal{A}}_v$ (right)
  with $A^{4(6)}_{\Delta \Gamma} = 0$.  The vertical bands correspond
  to $y_s$, with $\pm 1 \sigma$ (green) or $\pm 3 \sigma$ (yellow)
  errors. }
\label{AuAvpl}
\end{figure}

The relation between ${\cal{A}}_u$ and $[{\cal{A}}^{(2)}_T]_{theo}$ is
given in Eqs.~(\ref{Audef}) and (\ref{AT2rel1}); that between
${\cal{A}}_v$ and $[{\cal{A}}^{(1)}_T]_{theo}$ is given in
Eqs.~(\ref{Avdef}) and (\ref{AT1rel1}). These can be rewritten as
\beq
[{\cal{A}}^{(2)}_T]_{theo} \, \frac{\tau_{B_s}}{\langle\Gamma(\bspp)\rangle}
= -\frac{\pi}{2} {\cal{A}}_u \frac{(1 - y_s^2)}{(1 +
  A^{(4)}_{\Delta \Gamma} y_s)} ~, \nn
\eeq
\beq
[{\cal{A}}^{(1)}_T]_{theo} \, \frac{\tau_{B_s}}{\langle\Gamma(\bspp)\rangle}
= -\frac{\pi}{\sqrt{2}} {\cal{A}}_v \frac{(1 - y_s^2)}{(1 +
  A^{(6)}_{\Delta \Gamma} y_s)} ~.
\eeq
In Fig.~\ref{AuAvpl} we plot the dependence of the theoretical TP's
$[A^{(2)}_T]_{theo} \, {\tau_{B_s}}/{\langle\Gamma(\bspp)\rangle}$ and
$[A^{(1)}_T]_{theo} \, {\tau_{B_s}}/{\langle\Gamma(\bspp)\rangle}$ as a
function of $y_s$. The dashed black lines correspond to the central
values of ${\cal{A}}_u$ (left) and ${\cal{A}}_v$ (right) with
$A^{4(6)}_{\Delta \Gamma} = 0$.  The vertical bands correspond to
$y_s$, with $\pm 1 \sigma$ (green) or $\pm 3 \sigma$ (yellow) errors.
In the red and blue regions, we take $A^{4(6)}_{\Delta \Gamma} = \pm
1$, respectively, and allow ${\cal{A}}_u$ (left) and ${\cal{A}}_v$
(right) to vary by $\pm 1\sigma$ (see Table~\ref{LHCres}). It is clear
from these figures that, in the presence of NP, the values of the
theoretical TP's can differ significantly from the measured
asymmetries. (This is not surprising since the TP's vanish in the SM.)

\begin{figure}[t]
\begin{center}
\includegraphics[width=6cm]{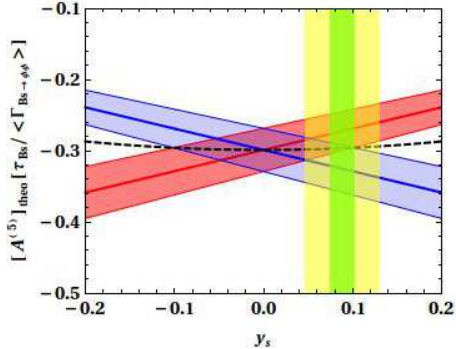}
\end{center}
\caption{The dependence of
  $[A^{(5)}_T]_{theo}{\tau_{B_s}}/{\langle\Gamma(\bspp)\rangle}$ on
  $y_s$ for different values of $A^{5}_{\Delta \Gamma}$.  In the red
  and blue regions, we take $A^{5}_{\Delta \Gamma} = \pm 1$,
  respectively. Also, $[A^{(5)}]_{exp}$ is allowed to vary by $\pm
  1\sigma$ [see Eq.~(\ref{A5value})]. The dashed black line
  corresponds to the central value of $[A^{(5)}]_{exp}$ with
  $A^{5}_{\Delta \Gamma} = 0$.  The vertical bands correspond to
  $y_s$, with $\pm 1 \sigma$ (green) or $\pm 3 \sigma$ (yellow)
  errors.}
\label{A5pl}
\end{figure}

Finally, for $i=5$, we have estimated the measured value of the
CP-conserving observable as follows:
\beq
\label{A5value}
[A^{(5)}]_{exp} = |A_0|_{exp} \, |A_\parallel|_{exp}
\, \cos(\delta_\parallel - \delta_0) = -0.299 \pm 0.030 ~.
\eeq
The relation between $[A^{(5)}]_{exp}$ and $[A^{(5)}]_{theo}$ is given
by [see Eq.~(\ref{A5rel})]
\beq
[A^{(5)}]_{theo} \, \frac{\tau_{B_s}}{\langle\Gamma(\bspp)\rangle}
= [A^{(5)}]_{exp} \, \frac{ (1 - y_s^2)}{(1 +
  A^{(5)}_{\Delta \Gamma} y_s)} ~.
\eeq
In Fig.~\ref{A5pl} we plot the dependence of
$[A^{(5)}]_{theo}{\tau_{B_s}}/{\langle\Gamma(\bspp)\rangle}$ as a
function of $y_s$. As before, the value of this quantity can differ
from Eq.~(\ref{A5value}) by as much as $O(10\%)$ for the current value
of $y_s$.

\section{Conclusions}

The main goal of studying the $B$ system is to find evidence for
physics beyond the standard model (SM). One possibility is new physics
(NP) in $\btos$ transitions. At present its status is uncertain.  It
seems unlikely that the effect of such NP can be very large, but a
smaller effect is still possible. In this paper, we consider $\btos$
NP. However, in contrast to what is usually done, i.e.\ considering
only NP in $\bs$-$\bsbar$ mixing, here we also allow NP in the
decay. In particular, we examine the effect of such NP on the angular
distribution of $B^0_q \to V_1 V_2$ ($q=d,s$), where $V_{1,2}$ are
vector mesons.

Our principal result is the following. The parameters of the untagged,
time-integrated angular distribution can be measured experimentally,
and certain observables can be derived from these parameters. However,
in the presence of NP, the formulae which relate the parameters to the
observables must be modified from their SM forms. We find six
observables for which the relation between the experimental data and
theoretical parameters must be modified, corresponding to the six
terms ($i=1$-6) in the angular distribution.  For $i=1$-3 they are the
polarization fractions, for $i=4$,6 they are the CP-violating
triple-product asymmetries, and $i=5$ corresponds to a CP-conserving
observable.  The modifications for the polarization fractions are most
interesting.  These are due in part to the nonzero width difference in
the $B^0_q$-${\bar B}^0_q$ system, and so are important only for $\bs$
decays. In particular, there can be important effects on the pure
$\btos$ penguin decay $\bspp$.

In light of this, we re-analyze the $\bspp$ data to see the effect of
these modifications. $\Delta \Gamma_s/2 \Gamma_s \sim 10\%$, so that
the modifications of the formulae lead to $O(10\%)$ changes in the
polarization fractions. These are not large, but may be important
given that one is looking for signals of NP.

Finally, if the NP contributes to the $\btos$ decay, we show that the
measurement of the untagged time-dependent angular distribution
provides enough information -- 12 observables -- to extract all the
NP parameters.

\bigskip
Note added: while the paper was being written, D\O\ produced a direct
measurement of the semileptonic charge asymmetry in $\bs$ decays
\cite{D0Bsasym}, and they say that it agrees with the SM. Technically,
this is true. The SM predicts $a^s_{sl} \sim 2 \times 10^{-5}$, and
D\O\ measures $a^s_{sl} = [-1.08 \pm 0.72~({\rm stat}) \pm 0.17~({\rm
    syst})]\%$. While this result is consistent with 0, the errors are
large enough that NP is also a possibility.

\section*{Acknowledgements} 
This work was financially supported by the US-Egypt Joint Board on
Scientific and Technological Co-operation award (Project ID: 1855)
administered by the US Department of Agriculture and in part by the
National Science Foundation under Grant No.\ NSF PHY-1068052 (AD and
MD), and by NSERC of Canada (DL).

\end{document}